%% file: manuscript_clean.tex
\documentclass[preprint, 3p, twocolumn]{elsarticle}

\usepackage{nomencl}
\usepackage{url}
\usepackage{hyperref}
\usepackage{gensymb}
\usepackage{xcolor}
\usepackage{array}
\usepackage{multirow}
\usepackage{amsmath}
\usepackage{enumitem}
\newcommand{\BHEX}{\emph{BHEX} }
\newcommand{\THEZA}{\emph{THEZA} }
\usepackage{orcidlink}

\usepackage{txfonts}
\usepackage{tikz}
\usetikzlibrary{shapes.geometric, arrows}

\tikzstyle{startstop} = [rectangle, rounded corners, 
minimum width=3cm, 
minimum height=1cm,
text centered, 
draw=black, 
fill=red!30]

\tikzstyle{io} = [trapezium, 
trapezium stretches=true, % A later addition
trapezium left angle=70, 
trapezium right angle=110, 
minimum width=2cm, 
text width=3cm,
minimum height=1cm, text centered, 
draw=black, fill=blue!30]

\tikzstyle{process} = [rectangle, 
minimum width=3cm, 
minimum height=1cm, 
text centered, 
text width=3cm, 
draw=black, 
fill=orange!30]

\tikzstyle{decision} = [diamond, 
minimum width=3cm, 
minimum height=1cm, 
text centered, 
text width=1.5cm, 
draw=black, 
fill=green!30]
\tikzstyle{arrow} = [thick,->,>=stealth]

\journal{Acta Astronautica}

\begin{document}

\begin{frontmatter}



\title{Toward Optimisation of a Sub-Terahertz Spaceborne VLBI Mission}


\author[inst1,inst2]{Ben~Hudson\corref{cor1}\orcidlink{0000-0002-3368-1864}}
\ead{benhudson@tudelft.nl}
\author[inst1,inst3]{Leonid~I.~Gurvits\orcidlink{0000-0002-0694-2459}}
\author[inst4,inst5]{Daniel~Palumbo\orcidlink{0000-0002-7179-3816}}
\author[inst4,inst6]{Sara~Issaoun\orcidlink{0000-0002-5297-921X}}
\author[inst4,inst5]{Hannah~Rana\orcidlink{0000-0001-6710-3387}}

\cortext[cor1]{Corresponding author}

\affiliation[inst1]{organization={Faculty of Aerospace Engineering, Delft University of Technology},%Department and Organization
            city={Delft},
            country={The Netherlands}}

\affiliation[inst2]{organization={KISPE Limited},%Department and Organization
            city={Farnborough},
            country={United Kingdom}}

\affiliation[inst3]{organization={Joint Institute for VLBI ERIC},%Department and Organization
            city={Dwingeloo},
            country={The Netherlands}
}

\affiliation[inst4]{organization={Center for Astrophysics $\vert$ Harvard \& Smithsonian},%Department and Organization
            city={Cambridge},
            country={USA}}

\affiliation[inst5]{organization={Black Hole Initiative},%Department and Organization
            city={Cambridge},
            country={USA}}

\affiliation[inst6]{organization={NASA Hubble Fellowship Program, Einstein Fellow},%Department and Organization
            country={USA}}

\begin{abstract}
Very Long Baseline Interferometry (VLBI) provides the finest angular resolution of all astronomical observation techniques. 
%The Earth-based Event Horizon Telescope (EHT) has demonstrated this in recent years with the landmark achievement of resolving the shadows of the supermassive black holes M87* and Sgr\,A*. However, these observations also showed that the science case for further sharpening the resolution of astrophysical studies is far from being exhausted.
However, observations with Earth-based instruments are approaching fundamental limits on angular resolution. These can only be overcome by placing at least one interferometric element in space.
%%The only way to overcome fundamental limits on angular resolution of Earth-based instruments is to place at least one interferometric element in space. 
In this paper, several concepts of spaceborne VLBI systems are discussed, including TeraHertz Exploration and Zooming-in for Astrophysics (\emph{THEZA}) and the Black Hole Explorer (\emph{BHEX}). Spaceborne VLBI telescopes have some of the most demanding requirements of any space science mission. The VLBI system as a whole includes globally distributed elements, each with their own functional constraints, limiting when observations can be performed. This necessitates optimisation of the system parameters in order to maximise the scientific return of the mission. 
%End-to-end mission simulations are an indispensable tool in conducting such an optimisation. 
Presented is an investigation into how the impact of the functional constraints of a spaceborne VLBI telescope affect the overall system performance. A preliminary analysis of how these constraints can be minimised through optimisation of the spacecraft configuration and operation is also provided. A space-based VLBI simulation tool (\texttt{spacevlbi}) has been developed to model such missions and its capabilities are demonstrated throughout the paper. It is imperative that the functional constraints are considered early in the design of the future space-based VLBI systems in order to generate feasible mission concepts and to identify the key technology developments required to mitigate these limitations.
\end{abstract}




\end{frontmatter}

\input{acronyms}

\section{Introduction}
\label{s:intro}
\noindent Very long baseline interferometry (VLBI) enables radio astronomers to study cosmic radio sources with the highest angular resolution of all astronomical techniques. Nowadays, VLBI  resolution reaches tens of microseconds of arc. Recently, the Event Horizon Telescope (EHT) achieved a major milestone by obtaining images of shadows of Supermassive Black Holes (SMBHs) in nuclei of the giant elliptical galaxy M87 (the radio source M87*) and our own Milky Way galaxy (the radio source Sgr\,A*). These observations achieved a resolution of \(\sim\)20~\(\mu\)as at a frequency of 230~GHz \cite{collaboration_first_2019, collaboration_first_2022}.  However, finer angular resolution is required for further advancements of our understanding of the physics of black holes as well as many other astrophysical phenomena.

The diffraction-limited angular resolution is characterised by a ratio of the observing wavelength $\lambda$ to the diameter or baseline length of the telescope $D$, $\lambda/D$.  Ground-based VLBI is inherently limited in angular resolution as the maximum baseline cannot exceed the diameter of the Earth and observational frequency is effectively constrained to below \(\sim\)350~GHz, due to atmospheric absorption. The EHT has already performed observations near this limit at 345~GHz and the next generation EHT (ngEHT) will regularly use this frequency \cite{roelofs_ngeht_2023,doeleman_reference_2023,raymond_first_2024}. Therefore, observation at as short a wavelength as possible and on longer baselines is essential for progressing the capabilities of VLBI. This can only be achieved by creating a spaceborne VLBI system which is not constrained by the Earth diameter and atmospheric limitations.

Two dedicated space VLBI missions have been flown as of the time of writing: VSOP-HALCA (1997--2003) and RadioAstron (2011--2019)  \cite[][and references therein]{LIG-2023-HISTELCON}. Following the recent successes of the EHT, multiple concepts are being proposed for the next generation of space-based VLBI systems. In this paper, two particular concepts are focused on: TeraHertz Exploration and Zooming-in for Astrophysics (\emph{THEZA}) \cite{gurvits_theza_2021}, and the Black Hole Explorer (\emph{BHEX}) \cite{BHEX-2024-SPIE}.

VLBI is a difficult technique to perform in space, requiring a complicated mission architecture and spacecraft design. As such, a number of functional constraints related to the system design limit when observations can be performed. This has the potential to severely reduce the science return of the mission. It is therefore essential that these constraints are identified early in the design process and are mitigated as much as possible. Presented in this paper is an investigation into the functional constraints pertaining to \emph{BHEX}, \THEZA and space-based VLBI in general, thus creating a starting point for optimisation of the mission architecture.

Section \ref{ss:VLBI} provides a brief overview of the VLBI technique. In section \ref{ss:Science}, the science objectives of the next generation space VLBI systems are discussed. Section \ref{s:spaceVLBI} describes the past space VLBI missions and the proposed concepts for the future space VLBI systems that appear in literature. Sections \ref{ss:THEZA} and \ref{ss:BHEX} provide more detailed descriptions of the \BHEX and \THEZA concepts. In section \ref{s:orbit}, the orbit configurations considered for \BHEX and \THEZA in this investigation are presented, along with rationale for their selection. In section \ref{s:optimisation}, an optimisation approach for the mission design is presented to minimise the impact of the functional constraints. Section \ref{s:constraints} includes identification of the major functional constraints impacting space-based VLBI and their specific relevance to \BHEX and \emph{THEZA}. Using the approach defined in section \ref{s:optimisation}, preliminary optimisation of the spacecraft design and mission architecture is performed to analyse mitigation strategies for each functional constraint. Finally, in section \ref{s:scienceReturn}, more general considerations for mitigating the impact on the science return of the future space VLBI missions are discussed.

\subsection{VLBI Basics}
\label{ss:VLBI}

\noindent In this subsection we give a very brief overview of VLBI relevant to the main subject of the current work. 

VLBI uses the principles of radio interferometry to enable signals received at spatially separated antenna to be correlated to estimate the ``visibility function'' of a source, corresponding to a Fourier transform of the sky image. These Fourier data carry the information on the true brightness distribution via various imaging methods. The angular resolution of such a system is dependent on the wavelength of observations and the projection on the image [or sky] plane of the baseline vector connecting the two antennas.

VLBI enables the separation between these antenna to be very large, hence providing very fine resolution, with no physical connection between the two antenna systems required. Observations are conducted concurrently at different facilities with a precise local frequency reference (typically an atomic clock) and time synchronisation between antennas. Data are stored on large solid state drives which are then transferred to a centralised location to undergo correlation.

A two-element interferometer provides a very sparse coverage of the source's visibility function, essentially a Fourier image of the brightness distribution in the observed source. The measurements are obtained at spatial frequencies $u$ and $v$, the independent variables in the Fourier domain which correspond to the source image plane rectangular coordinates. The set of \emph{(u,v)} points at which measurements of the visibility function are obtained is called the \emph{(u,v)} coverage. Inclusion of more antenna in the interferometer array, with a variety of baseline lengths throughout the observation, results in a denser \emph{(u,v)} coverage which is required for higher-fidelity reconstructions of the source image model.

\subsection{Space-based VLBI Science Objectives}
\label{ss:Science}
\noindent Space-based VLBI can enable scientific investigations fundamentally impossible with ground-based VLBI facilities. The scientific objectives of the next generation of space VLBI systems are wide ranging, with the potential to answer key questions across multiple areas of research. One of these areas is the phenomenon of photon rings around black holes -- the primary objective of \BHEX \cite{BHEX-2024-SPIE} and one of the major scientific tasks of \THEZA \cite{gurvits_theza_2021, gurvits_science_2022, Hudson_2023}.

Background photons in the vicinity of a SMBH can experience extreme deflection and complete \(n\) half-orbits about the black hole before escaping. The effect represents an ultimate case of gravitational lensing. As \(n\) increases, the observer sees an exponentially sharper, but less luminous in terms of total flux density, feature \cite{johnson_universal_2020}. Successive subrings asymptotically approach the boundary of the black hole shadow. As the photon ring approaches this critical curve, its structure depends increasingly on spacetime geometry and less on astrophysical phenomena \cite{gralla_black_2019}. However, even at low values of \(n\), the photon ring geometry contains information on the black hole's mass and spin, and can be used to conduct strong-field tests of general relativity (GR) \cite{johnson_universal_2020, gralla_shape_2020, broderick_measuring_2022, wielgus_photon_2021}. Spin can also be constrained through other methods. Palumbo et al. shows a correlation between spin and the curl of the linear polarization pattern in the emission ring in GRMHD simulations \cite{palumbo_discriminating_2020,chael_black_2023} This can be performed on sources that cannot be resolved at horizon-scale like M87* and Sgr\,A*, enabling \BHEX to unlock a range of SMBH for which such measurements may be possible.

On Earth baselines at 230~GHz and 345~GHz, the EHT and ngEHT cannot yet resolve the \(n=1\) ring \cite{collaboration_first_2022,johnson_key_2023}. Johnson et al. demonstrate that the interferometric signature of a black hole on very long baselines is dominated by the photon ring contribution \cite{johnson_universal_2020}. As can be seen in Fig. 4 of Johnson et al., observation on baselines longer than \(\sim\)20~G\(\lambda\) provides access to the photon ring-dominated regime \cite{johnson_universal_2020}. Therefore, if an interferometer can observe on these long baselines with sufficient sensitivity to detect the decreasingly weak signals, the photon rings can be characterised \cite{gralla_shape_2020, paugnat_photon_2022}. This will be possible with space-based VLBI systems and detection of the \(n=1\) ring is the primary objective of \BHEX \cite{BHEX-2024-SPIE}. Probing \(n=2\) with future space VLBI missions would provide the most accurate tests of strong gravity to date \cite{gralla_shape_2020, paugnat_photon_2022}.

Prospective Space VLBI science is not only focused on photon ring detection. Earth-based arrays cannot yet capture the dynamic behaviour of M87* or Sgr\,A*, with gravitational timescales of \(\sim\)9~hours and 20~seconds, respectively \cite{johnson_key_2023}. Longer observations are required to study the evolution of M87* and Earth-rotation does not provide rapid enough variation in the \emph{(u,v)} plane to capture the dynamic behaviour of Sgr\,A*. Space-based systems can achieve rapid filling of the \emph{(u,v)} plane and may even enable the generation of movies of black hole evolution, from multiple image reconstructions of the source. 

Space VLBI will also enable study of other AGN and their jets, providing insight into the bright, compact core feature seen in many blazars \cite{marscher_inner_2008}. This will enable study of jet acceleration and collimation, providing a means of constraining possible energy extraction mechanisms and jet models \cite{kovalev_transition_2020, blandford_electromagnetic_1977, tursunov_fifty_2019}.

An intriguing science case for future space-based VLBI systems is the study of sub-parsec, SMBH binary systems (SMBHB). SMBHBs are believed to be products of galaxy mergers and the dynamics of their inspiraling at a certain stage is dominated by gravitational wave (GW) emission \cite{colpi_massive_2014, armitage_accretion_2002, milosavljevic_final_2003}. A space VLBI mission providing sufficient angular resolution and sensitivity to resolve such SMBH binaries would enable direct measurement of the system's properties. As such, multi-messenger observations of these objects, in collaboration with instruments such as the spaceborne LISA and a variety of Earth-based GW instruments, would provide a rich study of the origin and evolution of black hole mergers \cite{PhysRevD.106.103017}.

In this paper, photon ring detection and the subsequent investigations that can be performed with characterisation of the ring properties are considered as the primary objective of the space VLBI concepts under analysis. M87* and Sgr\,A* are considered as the target sources although discussion is given to the increase in mission optimisation complexity if other sources are considered as well. Future work will investigate mission design optimisation for performing more specific AGN studies and multi-messenger observations of black hole binaries.

\vskip 1em
\section{Space-Based VLBI}
\label{s:spaceVLBI}
\noindent As stated in section \ref{s:intro}, expansion of VLBI arrays into space is crucial to overcome the limitation on angular resolutions of Earth-based arrays. However, inherent to space VLBI design and functionality  are a range of constraints which limit observations. These constraints differ from and are more diverse than the well accustomed observational limitations of ground-based VLBI (weather, local horizon masks, sky coverage, etc.). The need of addressing functional limitations was understood and addressed at an early stage of the RadioAstron mission development \cite{LIG-1991-Sagami} and later adopted, at least partially, in both dedicated space VLBI missions flown by the time of writing, VSOP-HALCA and RadioAstron.

VSOP-HALCA utilised an 8~m antenna and operated in a Highly Elliptical Orbit (HEO), with an apogee of \(\sim\)21,400~km and an orbital period of 6.3~hours \cite{Hirax+1998Sci}. The spacecraft required a radio link with a tracking ground station to be maintained throughout observations. However, unlike RadioAstron, VSOP-HALCA did not fly a frequency standard onboard. Instead, the link with the ground station was used to transmit a high-stability signal to feed into the onboard heterodynes. Across all precessions of its highly elliptical orbit, this resulted in a range of 0.55 and 0.85 fractions of the orbit when a ground station was available, greatly reducing the duration of potential observation periods \cite{hirabayashi_vlbi_2000}. 

RadioAstron utilised a 10~m, deployable antenna and also operated in a Highly Elliptical Orbit (HEO) but with an apogee near lunar in distance \cite{kardashev_radioastron-telescope_2013}. On the longest baselines and highest frequencies, it achieved the finest angular resolution in continuum imaging observations, 27~\(\mu\)as \cite{Fuentes+2023NatAs}. The RadioAstron User Manual provides a clear description of the observational constraints of the system and the reader is referred to that document for more detail \cite{radioastron_science_and_technical_operations_group_radioastron_2019}. For example, significant constraints existed in the thermal control of the spacecraft with a very limited range of Sun positions for which observations could be performed. For both VSOP-HALCA and RadioAstron, tracking from ground stations was also required to provide accurate position determination of the spacecraft which is crucial for the correlation process post observations \cite{murata_vsophalca_2000,radioastron_science_and_technical_operations_group_radioastron_2019}.

\begin{figure}[t!]
    \centering
    \includegraphics[width=\columnwidth]{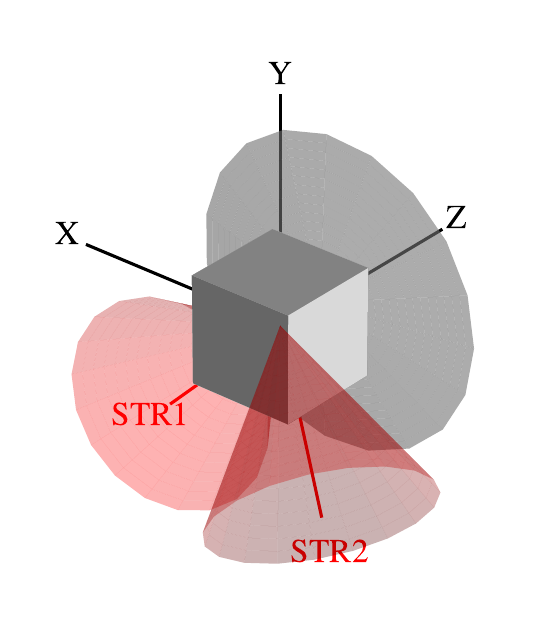}
    \caption{Simplified, example configuration of a space-based VLBI system, illustrating body-fixed coordinate system, antenna position and example star tracker (STR) locations.}
    \label{f:spacecraft}   
\end{figure}

It is essential that lessons are learnt from the experience of VSOP-HALCA and RadioAstron in order to maximise the science return of future space VLBI missions \cite{gurvits_space_2020}. In Fig. \ref{f:spacecraft} is depicted an example, highly simplified configuration of a space-based VLBI system to provide clarity in the following discussions which shall refer to specific component positions with respect to the body-fixed spacecraft cooordinate system.

\subsection{Future Mission Concepts}
\label{ss:concepts}
\noindent Several, space-based VLBI concepts have been proposed in recent years, of varying levels of maturity. The mm/sub-mm Millimetron mission has been proposed as a followup of the RadioAstron mission \cite{NSK+1995AcAau, NSK+2014PhyU}. The Event Horizon Imager (EHI) concept has also been discussed \cite{roelofs_simulations_2019, shlentsova_imaging_2024}. The EHI consists of two to three space telescopes in Medium Earth Orbit (MEO), aiming to achieve a resolution of \(\sim\)5~\(\mu\)as with space-space baselines. The  CAPELLA concept proposes two pairs of small VLBI satellites operating in polar, Low Earth Orbits (LEO) and observing at 690~GHz to provide a resolution of 7~\(\mu\)as \cite{CAPELLA-2023arXiv}. Forming ground-space baselines, Fish et al. propose two spacecraft operating in a Geostationary Earth Orbit (GEO) and a HEO as an addition to the EHT, to increase resolution to 3~\(\mu\)as \cite{Fish+2020}.

Although some of these concepts have provided commentary on the practical engineering difficulties associated with space-based VLBI, none have as of yet considered the detailed operational impact of functional constraints. All of the referenced concepts discuss optimising the spacecraft's orbit to meet the science objectives but leave aside the functional constraints as a key element of the overall design of the mission architecture and instrumentation. This is an essential activity, particularly when such concepts mature to the point of considering the design of the system, as the science return of the mission may be severely degraded if these issues are not tackled early in the development. 

The two mission concepts that are considered in detail in this paper are described more fully below.

\subsection{THEZA}
\label{ss:THEZA}
\noindent TeraHertz Exploration and Zoming-in for Astrophysics (\emph{THEZA}) is a concept that was prepared in response to the European Space Agency's (ESA) call for its next science program Voyage~2050 \cite{gurvits_theza_2021,gurvits_science_2022}. By observing at millimeter and sub-millimeter wavelengths, \THEZA aims to achieve at least an order of magnitude improvement on the EHT's angular resolution. The \THEZA concept consists of at least two spaceborne antennas, forming space-space baselines.

Although \THEZA has not undergone a detailed mission design exercise, key elements of the system have been discussed \cite{gurvits_theza_2021, gurvits_science_2022, Hudson_2023}. Two methodologies are presented for the handling of the large volumes of VLBI data for \emph{THEZA}. As with \emph{BHEX}, the use of optical communications systems onboard is proposed to downlink the data to the ground. An alternative architecture consisting of in-orbit correlation by transferring the data from one spacecraft to the other via optical Inter-Satellite Link (ISL) is also discussed \cite{gurvits_science_2022}. The latter being a more forward-looking approach that would require significant computing capability onboard. As such, in this paper, downlink of science data to the ground via optical terminal to a set of dedicated, optically-capable ground stations is considered in the evaluation of the impact of the functional constraints.

\subsection{BHEX}
\label{ss:BHEX}
\noindent The Black Hole Explorer (\emph{BHEX})\footnote{\url{https://www.blackholeexplorer.org/}} is a concept aiming at a NASA Small Explorer (SMEX) programme at the next call for proposals expected in 2025 \cite{BHEX-2024-SPIE}. A key science aim of \BHEX is to resolve the photon rings of M87* and Sgr\,A* and study fundamental properties of the black holes directly, such as mass and spin. This will be accomplished with a 5x increase in angular resolution compared to the EHT.

The preliminary \BHEX design \cite{Peretz+SPIE-2024} consists of a 3.5~m, fixed, monolithic antenna and cryogenic cooling of the receiver system to achieve sufficient Signal-Noise Ratio (SNR) to enable detections of the \(n=1\) photon ring. Observing across two frequency bands in collaboration with ground-based antenna, the primary receiver will be double-side-band (DSB) and operate over a frequency 240--320~GHz. A secondary receiver will be single-side-band (SSB), operating over the range 80--106~GHz \cite{BHEX-2024-SPIE}. Both receivers will be capable of dual-polarization measurements. \emph{BHEX}'s maximum observational frequency of 320~GHz is used in the subsequent generation of \emph{(u,v)} coverage plots. \BHEX will utilise a real-time, optical downlink of raw VLBI data. This system will be based on that demonstrated by NASA's TBIRD mission, achieving 100~Gbps from the reference \BHEX orbit \cite{Wang+2023SPIE,BHEX-2024-SPIE}. A crystal ultra-stable oscillator (USO) will also be flown onboard the spacecraft to provide a stable frequency reference to achieve coherence during observations \cite{marrone_black_2024}.

In evaluating the impact of the functional constraints on \BHEX observations, the following onboard subsystems are considered: attitude control (e.g. blinding of star trackers), thermal (e.g. deep-space radiator pointing), communications and data handling (e.g. real-time, optical downlink) and power (e.g. solar panel power generation). 

\vskip 1em
\section{Mission Simulation: \texttt{spacevlbi}}
\label{s:python}

\noindent Simulation of space VLBI observations is a powerful tool for analysis and synthesis of a mission during its design stage, as well as an instrument supporting mission operations. Over the past half a century, various space VLBI simulation packages have been used for design studies of several dozen projects. Arguably, the most advanced and practically important was the package Fakesat developed at JPL \cite{Joel-Smith+2000}. This package was used during the development and in-orbit operations of both dedicated space VLBI missions flown to date, VSOP/HALCA and RadioAstron. However, this package was created using currently obsolete software and cannot be used for the needs of the present study.

In order to model space-based VLBI missions, the Python package \texttt{spacevlbi} has been developed \cite{hudson_python_2024}. \texttt{spacevlbi} is capable of high fidelity modelling of a space telescope's orbit and attitude state. Ground VLBI antennas can also be modelled, enabling ground-space VLBI simulations. By defining a ground array, a space telescope orbit and a target source, \texttt{spacevlbi} provides the \emph{(u,v)} coverage achieved over the simulation time period. Multiple space telescopes can also be modelled simultaneously, enabling simulation of missions such as \THEZA.

The attitude of the space telescope is modelled throughout the simulation by defining a body-fixed axis to point towards a target source. By modelling the attitude of the spacecraft, the directions in which key onboard instruments and subsystems such as the antenna, solar panels, optical terminals, star trackers and radiator surfaces are pointed can be determined. This allows the impact of the functional constraints associated with each of these units to be evaluated.

This will also make the tool of use during conceptual and detailed design phases of future space-based VLBI systems, allowing the practical considerations of spacecraft design to be combined with more detailed science simulations such as those that can be performed in packages like the \texttt{eht-imaging} library \footnote{\url{https://github.com/achael/eht-imaging}} \cite{chael_interferometric_2018}. The functional constraints that can be modelled in \texttt{spacevlbi} cover a wide range of traditional spacecraft subsystems and the reader is referred to the documentation of the publicly available tool for more information on its capabilities \footnote{\url{https://github.com/bhudson2/spacevlbi}}. 

\texttt{spacevlbi} is used to perform all of the analysis presented in this paper. \emph{(u,v)} coverage figures are based purely on the geometry of the interferometer and the target source and do not take into account effects which limit the observing ability of Earth-based radio telescopes, such as weather conditions and other temporary factors. For the purpose of demonstrating the impact of the mission functional constraints, characteristics of real VLBI observations such as the integration time, length of each scan and the cadence that scans are performed at have also not been considered in the analysis. The optimisation approach and analysis of missions such as \BHEX presented here is agnostic of the values of these parameters.

The various functionalities of \texttt{spacevlbi} have been verified against independent tools to ensure they are performing as expected. The orbit propagator makes extensive use of the \texttt{poliastro} package  \footnote{\url{https://github.com/poliastro}}. The orbit propagation has undergone extensive testing as described in the \textit{validation} repository accessible via the Github link. The \emph{(u,v)} coverage calculated by \texttt{spacevlbi} has been verified against that generated by \texttt{eht-imaging}. The results agree within the bounds of the known differences in propagation technique used by the simulators (\texttt{eht-imaging} assumes an unperturbed Keplerian orbit). Celestial data such as Sun and Moon positions is downloaded directly from the DE440 JPL Planetary Ephemeris files \cite{park_jpl_2021}.

\section{Orbit Configuration}
\label{s:orbit}
\noindent Orbit selection for science missions is typically a trade-off between what is optimal to meet the scientific objectives and feasibility from an engineering perspective. Preliminary orbits are being considered for \BHEX based on the scientific objectives of the mission and some consideration of the engineering implementation. However, analysis of the functional constraints will be required regardless of the final orbit selection and it is the methodology presented in this paper that is the focus.

\begin{figure}[t!]
    \centering
    \includegraphics[width=\columnwidth]{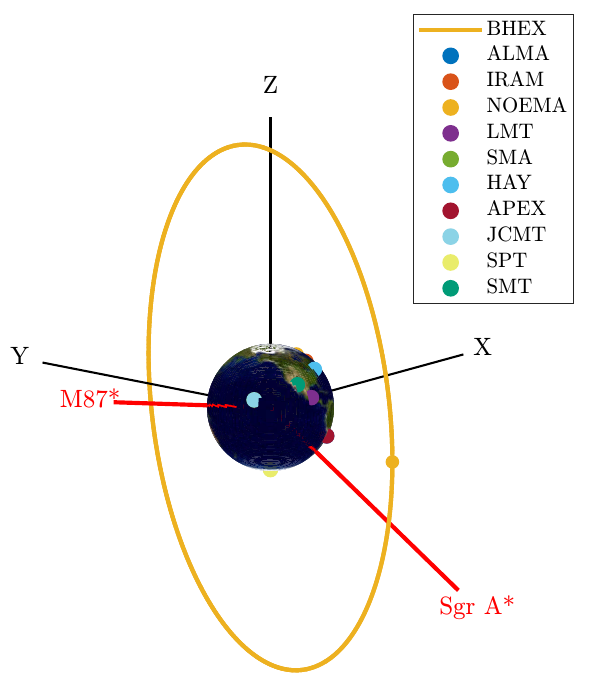}
    \caption{Preliminary \BHEX orbit for the detection of the \(n=1\) photon ring in M87* and Sgr\,A*.}
    \label{f:bhexOrbit}   
\end{figure}

\begin{figure}
    \centering
    \includegraphics[width=\columnwidth]{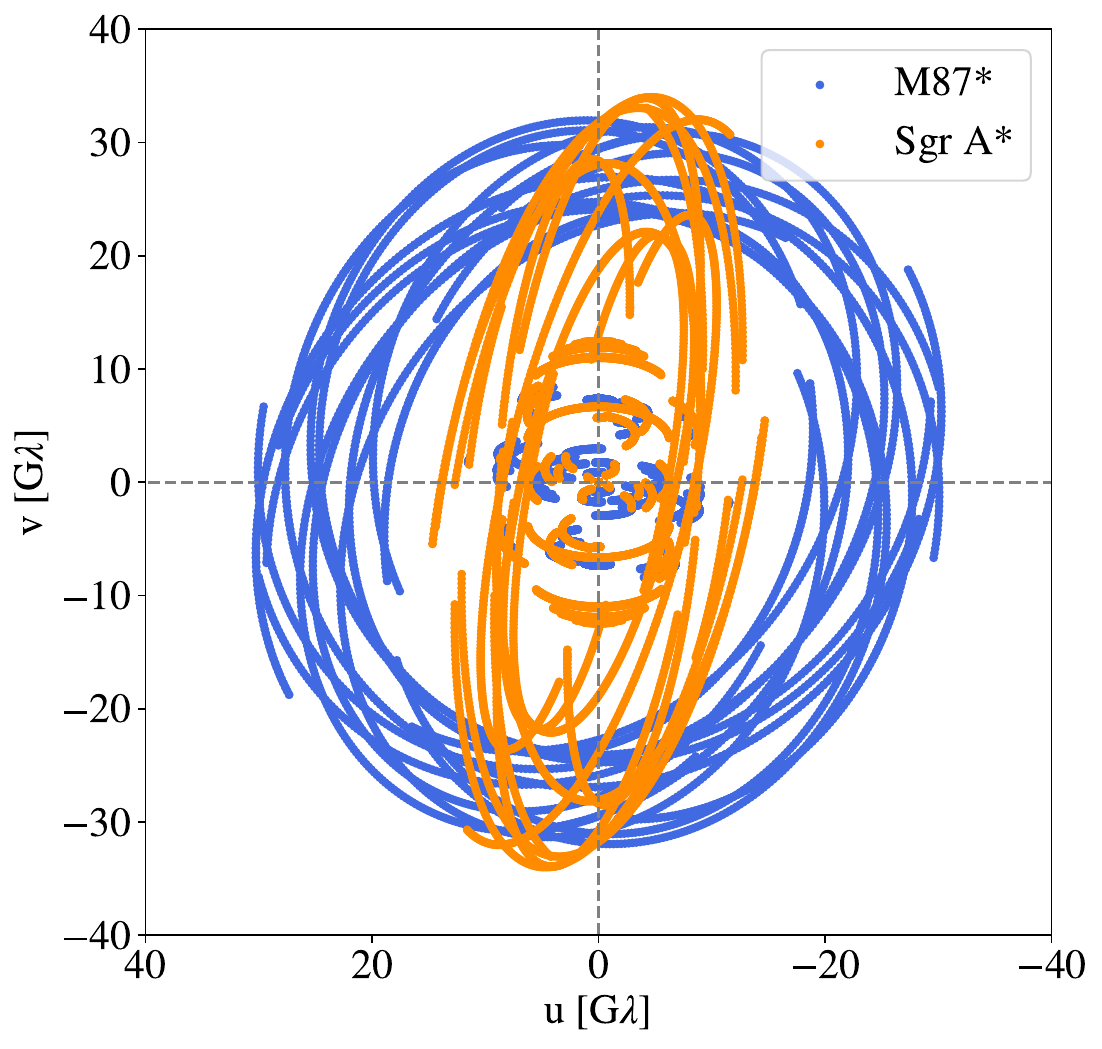}
    \caption{An ``ideal'' \emph{(u,v)} coverage for M87* and Sgr\,A* achieved by \BHEX in the reference orbit. Observation conducted with ground array depicted in Fig. \ref{f:bhexOrbit}, at a frequency of 320~GHz, over 24~hours.}
    \label{f:bhexUV}   
\end{figure}

The orbit selection for \BHEX is still taking place balancing the need to meet the science requirements and maintaining engineering feasibility. A favoured candidate for \BHEX is a circular, polar MEO with an orbital radius of 26562~km, and it is likely that the final selected orbit will be of this class \cite{BHEX-2024-SPIE}. This orbit has been preliminarily selected for the following reasons:
\begin{itemize}
    \item It provides the required space-ground baseline coverage (and hence angular resolution) of both sources for detection of the first order photon ring. The full baseline range offered by \BHEX together with the specified ground-based array is \(\sim\)18-33~G\(\lambda\) (M87*) and \(\sim\)2-34~G\(\lambda\) (Sgr\,A*), at 320~GHz \cite{johnson_universal_2020}.
    \item For M87*, precise estimates of the mass are not available which requires that mass/spin degeneracies in the photon ring must be broken with two-dimensional information about its shape and relative astrometry. This requires the near-circular \emph{(u,v)} coverage by selecting an orbital plane almost perpendicular to M87*. For Sgr\,A*, knowledge of the mass is available to  a finer degree of precision. This means that a photon ring size measurement along even a single axis provides an excellent spin constraint. The orbital plane has however been rotated slightly to increase the coverage of Sgr\,A* in the \emph{u}-plane.
    \item The orbital period of 12 sidereal hours generates a repeating ground-track which simplifies the maintenance of real-time downlink as the same ground locations will be able to perform a link session with \BHEX at the same time each day (see Fig. \ref{f:ground_track} for a depiction of this phenomena).
\end{itemize}
 
\noindent The unconstrained \emph{(u,v)} coverage achieved by \BHEX in collaboration with an extensive ground array (see Fig.~\ref{f:bhexOrbit} for definition of the ground array) is shown in Fig.~\ref{f:bhexUV}. A minimum observing elevation of 15\(\degree\) is assumed at each ground station.

In our previous work a number of orbit configurations are presented for a spaceborne, two-element \THEZA system, utilising 15~m antennas and optimised for the detection of photon rings \cite{Hudson_2023}. The configurations have been designed to achieve a wide variation in baseline length of 20-200~G\(\lambda\), irrespective of the position of the source on the sky.

\begin{figure}[t!]
    \centering
    \includegraphics[width=\columnwidth]{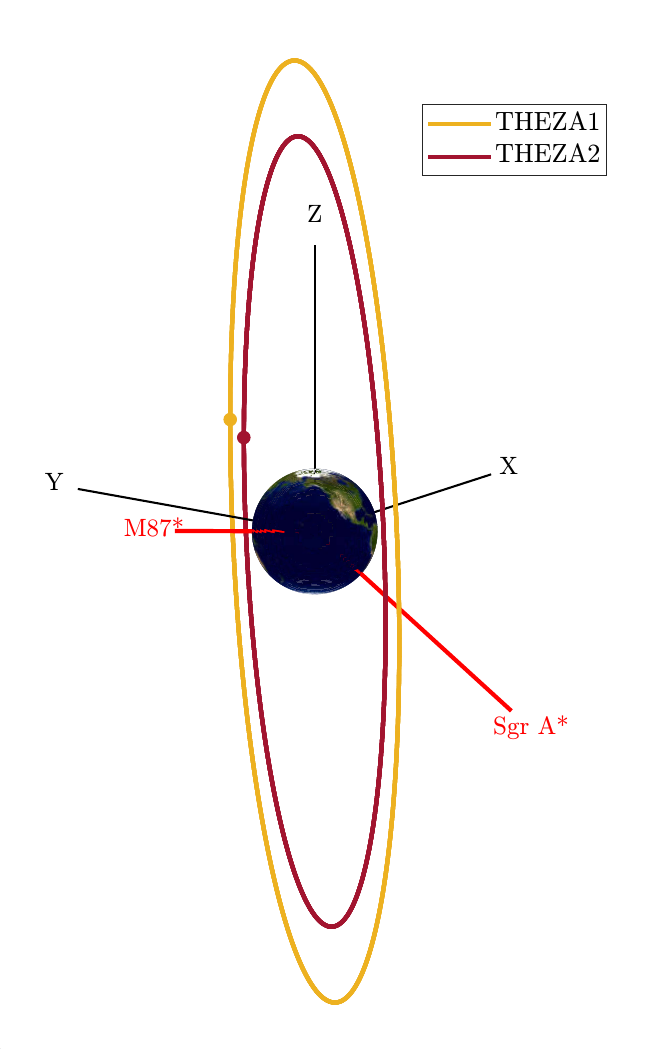}
    \caption{Circular, coplanar, polar orbit configuration of \THEZA, for the detection of the \(n=1\) and \(n=2\) at 690~GHz \cite{Hudson_2023}.}
    \label{f:thezaOrbit}   
\end{figure}

\begin{figure}
    \centering
    \includegraphics[width=\columnwidth]{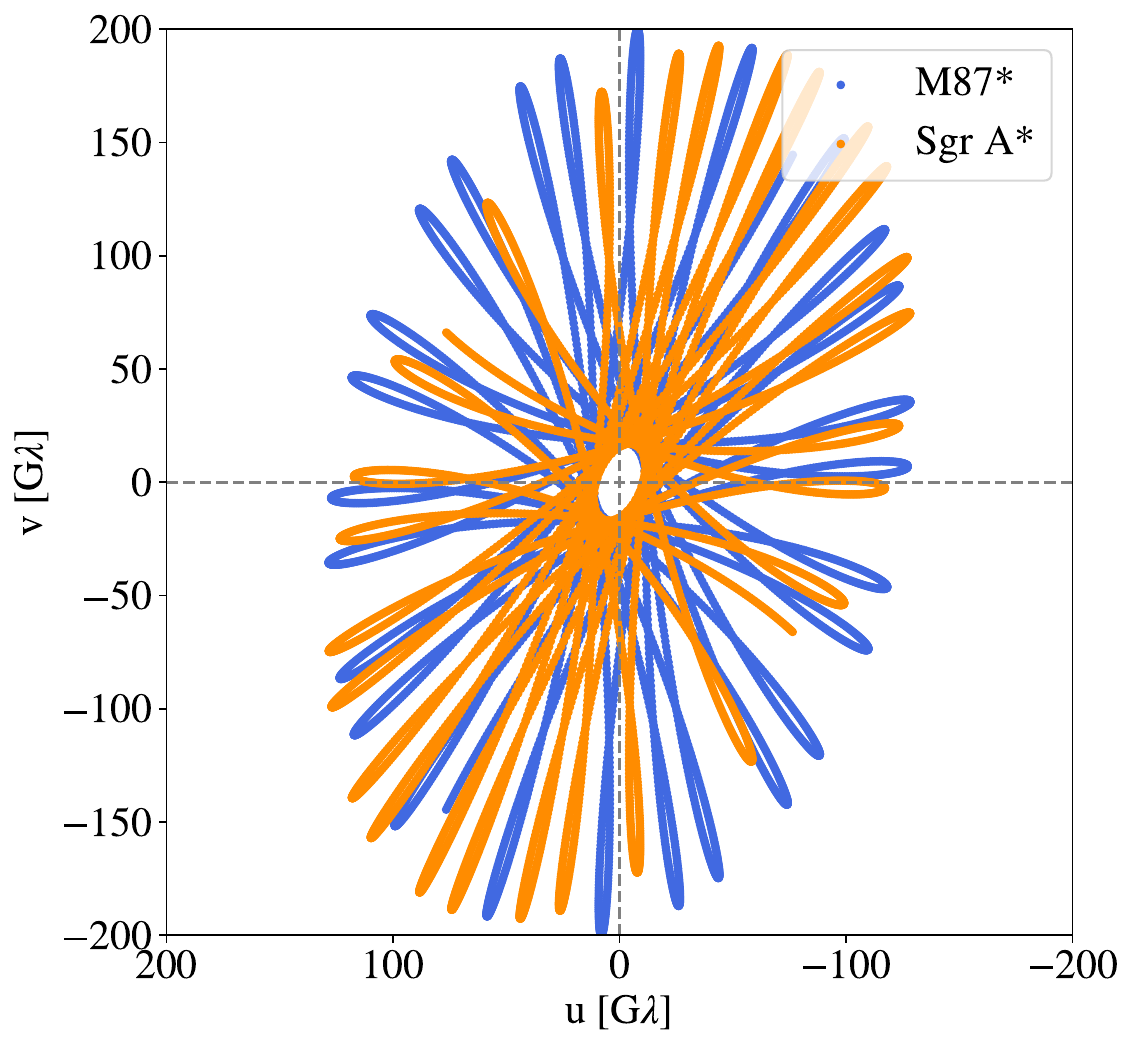}
    \caption{An ``ideal'' \emph{(u,v)} coverage for M87* and Sgr\,A* achieved by \THEZA in the orbit depicted in Fig. \ref{f:thezaOrbit}. Observation conducted at 690~GHz across 7~days.}
    \label{f:thezaUV}   
\end{figure}

Fig. \ref{f:thezaOrbit} depicts the circular, coplanar MEO configuration of \THEZA from \cite{Hudson_2023}. The right ascension of the ascending node of the orbits has been set to the average of M87* and Sgr\,A*, maximising \emph{(u,v)} coverage for both sources (depicted in Fig. \ref{f:thezaUV}). The starting true anomaly of either spacecraft can be set to any value and the required baseline variation will still be achieved within 7~days.

\vskip 1em
\noindent The reference \BHEX orbit and the \THEZA orbit depicted in Fig. \ref{f:thezaOrbit} are used as the example configurations of the two space VLBI systems in the subsequent functional constraint analyses.

\vskip 1em
\section{Mission Optimisation}
\label{s:optimisation}
\noindent Space mission development is an iterative process whereby a design slowly converges to a point where it meets all of the mission requirements. Trade-offs are an inherent part of spacecraft design and identification of the key trade spaces and resolution of these conflicts is a core part of the development process (i.e. the needs of one subsystem are often not compatible with those of another).

Although the design of a space VLBI mission will follow this same process, it is unique among space science applications in the complexity of its operation. Not only does it require a highly performant spacecraft, the mission architecture consists of multiple, complex relationships with systems on the ground (e.g., ground radio telescopes and tracking, command and control, and specialised data acquisition stations). As such, space-based VLBI poses a unique design study task, one that warrants a bespoke methodology in the optimisation of the system's characteristics.

Presented in this section is a configuration optimisation method, using \texttt{spacevlbi}, that enables the optimal position of spacecraft components with external Fields of View (FOV) to be determined. This method is useful for positioning items such as star trackers, radiator surfaces, communication systems and solar panels, or any other units whose performance is dependent on a specific relationship with the Sun, Earth or Moon. The proposed approach also evaluates the entire search space, providing the user with all possible, optimal configurations, given the constraints from the spacecraft model.

%This method has been developed in order to find an optimal spacecraft configuration to minimise the impact of the functional constraints on space VLBI observations. The optimisation is based upon simulations performed with the \texttt{spacevlbi} package (presented in section \ref{s:python}), which utilises the DE440 JPL Planetary Ephemeris files to provide Sun and Moon positions \cite{park_jpl_2021}. This enables highly accurate and time-dependent analyses of the impact of Sun, Earth and Moon position on the spacecraft operation.

\begin{figure}
    \centering
    \includegraphics[width=\columnwidth]{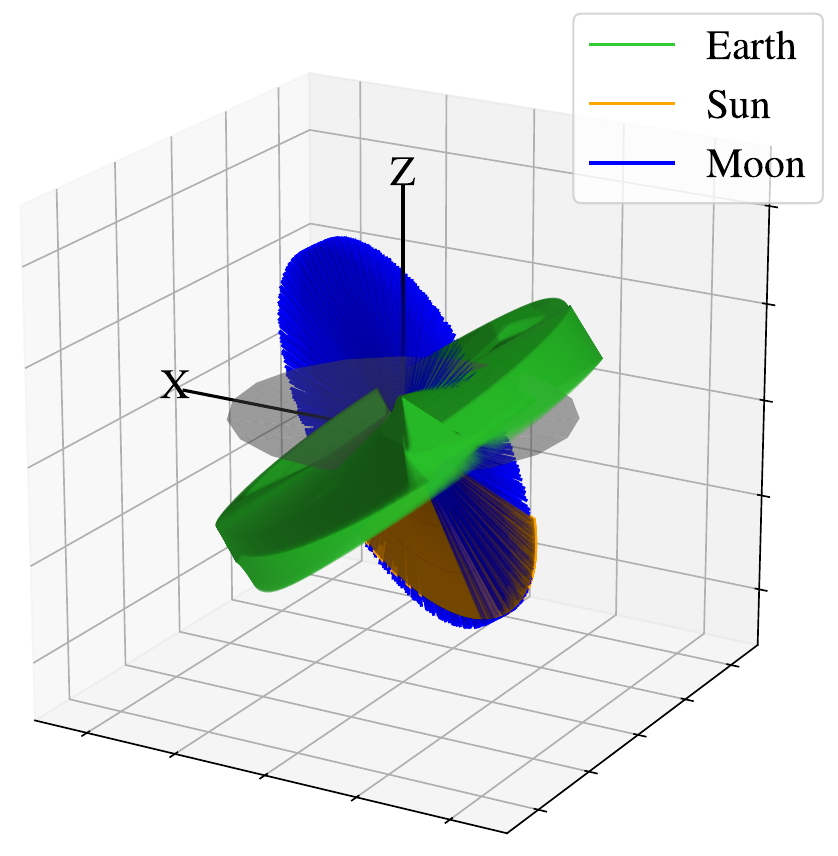}
    \caption{Attitude sphere showing \BHEX antenna configuration and Sun, Earth and Moon positions throughout Jan--Apr observation season of M87*.}
    \label{f:AttitudeSphere}   
\end{figure}

Use of the \emph{attitude sphere} is the core of the optimisation method. An attitude sphere depicts the spacecraft body-fixed axis, the pointing of various components and the positions of the Sun, Earth and Moon, over a given period of time. Fig. \ref{f:AttitudeSphere} depicts the attitude sphere of \BHEX and shows the Sun, Earth and Moon positions when observing M87* across the Jan--Apr observation season. The optimisation method is then performed as described in section \ref{s:A1}.

\vskip 1em
\section{Functional Constraints}
\label{s:constraints}
\noindent In this section, an investigation into the functional constraints that impact space VLBI missions is performed, along with an assessment of how these will limit the mission's observational capabilities. Specifically, the \BHEX concept is used as a case study. However, discussion of how the constraints are relevant for other concepts such as \THEZA is also included.

\begin{figure*}
    \centering
    \includegraphics[width=0.9\textwidth]{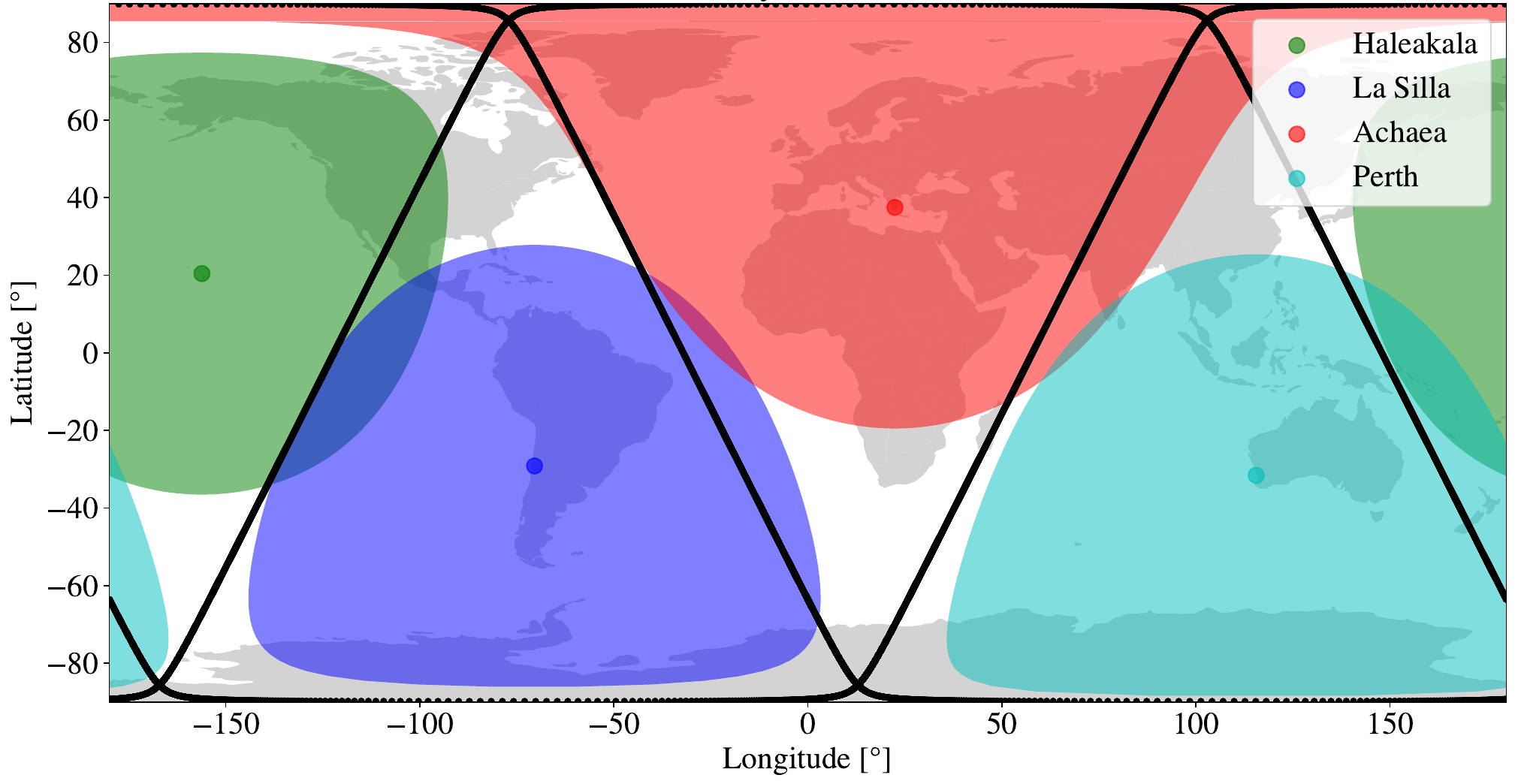}
    \caption{\BHEX ground track in preliminary orbit. Preliminary optical ground station locations depicted with visibility of \BHEX when minimum elevation is set to 20\(\degree\).}
    \label{f:ground_track}   
\end{figure*}

\subsection{Science Data Communications and Handling}
\label{ss:CDH}

\noindent Data handling of a space VLBI mission is a particularly challenging aspect of the design. VLBI requires operation at very wide bandwidths to increase sensitivity and therefore, data are recorded at extremely high rates. For a 24~hour observation, a single interferometer element might generate terabytes of data. Since 2018, the EHT has been operating at 64~Gbps \cite{collaboration_first_2019}. A trade-off exists between the implementation of mass data storage onboard and the real-time downlink of data to the ground. A description of this trade-off and the current state of the technology required for both options are discussed for \THEZA in \cite{gurvits_science_2022}.

The current concept of operations for \BHEX involves the use of a real-time downlink of data, using an optical communications terminal. This system is based on that demonstrated by the TBIRD mission, which has demonstrated downlink rates of up to 200~Gbps \cite{schieler_-orbit_2023, Wang+2023SPIE}. The optical communications
onboard terminal will be gimballed, allowing it to point independently (within certain limits) of the spacecraft's attitude. The limits of the gimbal capability are still under consideration, and the effect of varying this is presented in the subsequent analysis.

During observations, \BHEX must remain in an inertially-fixed attitude in order to point the antenna at the target source and minimise variation in the polarisation of the received signals. Therefore, the Earth-facing side of the spacecraft would vary throughout observations, as the spacecraft moves around its orbit. Under these restrictions, the optimal position for the unit is in the opposite direction to the antenna, with the nominal pointing direction along the negative Z-axis. This was calculated with the process described in section \ref{s:optimisation}. With the gimbal capability, the terminal can then point in almost any direction in the negative Z-hemisphere. Observation of either source can only take place if an optical link can be maintained with at least one ground station of an array distributed across the globe. A minimum elevation at each ground station of 20\(\degree\) is assumed. For this analysis, a set of four ground optical sites distributed across the globe are considered: Perth Australia, Haleakala Hawaii, Achaea Greece, La Silla Chile. These sites have been preliminary selected for BHEX as they provide constant coverage of the orbit. This mission architecture is shown in Fig. \ref{f:ground_track}.

This configuration results in a loss of 44\% and 49\% of \emph{(u,v)} coverage when observing M87* and Sgr\,A*, respectively. This undesirable impact on observations could be mitigated through a number of methods:
\begin{itemize}[noitemsep]
    \item{Inclusion of mass data storage for onboard buffering of VLBI data during periods of the orbit when a link with the ground cannot be achieved with subsequent data downlink sessions that are asynchronous with observations;}
    \item{Multiple optical communication terminals located around the spacecraft bus;}
    \item{Position of the optical terminal and rotation of the spacecraft about the antenna direction (Z-axis), at intermittent points during observations, to keep the Earth within the FOV of the gimballed terminal.}
\end{itemize}

\begin{figure}
    \centering
    \includegraphics[width=\columnwidth]{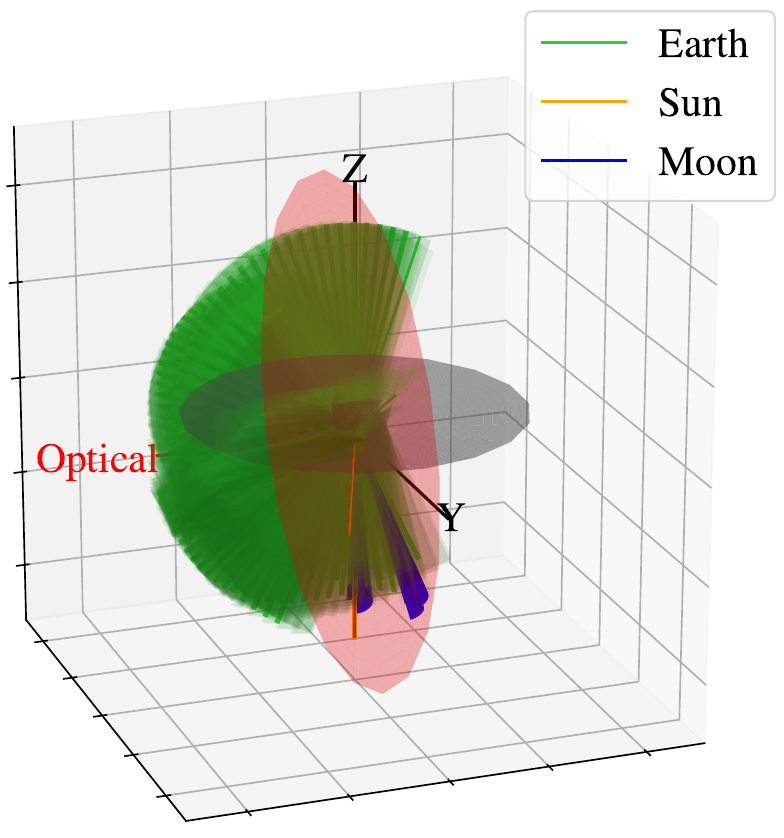}
    \caption{Attitude sphere showing optimal \BHEX optical terminal configuration and Sun, Earth and Moon positions during observation of M87* and Sgr\,A* (1\textsuperscript{st} Jan and 1\textsuperscript{st} June, respectively.).}
    \label{f:CDHsphere}   
\end{figure}

\noindent The last solution is the most feasible, given the programmatic constraints of the \BHEX project. Fig. \ref{f:CDHsphere} depicts an optical terminal configuration that provides an almost constant real-time downlink with ground stations, when observing M87* and Sgr\,A*. The terminal is mounted in the spacecraft +X axis and every half orbit period, the spacecraft is rotated 180\(\degree\) about the +Z axis to keep the Earth within the optical terminal FOV. By performing this rotation, the variation in polarisation of the received signals is negated. This strategy also provides a deep-space facing side of the spacecraft that would be highly beneficial for the mounting of star trackers and thermal components. This is discussed in more detail in the following sections.

\begin{figure}[t!]
    \centering
    \includegraphics[width=\columnwidth]{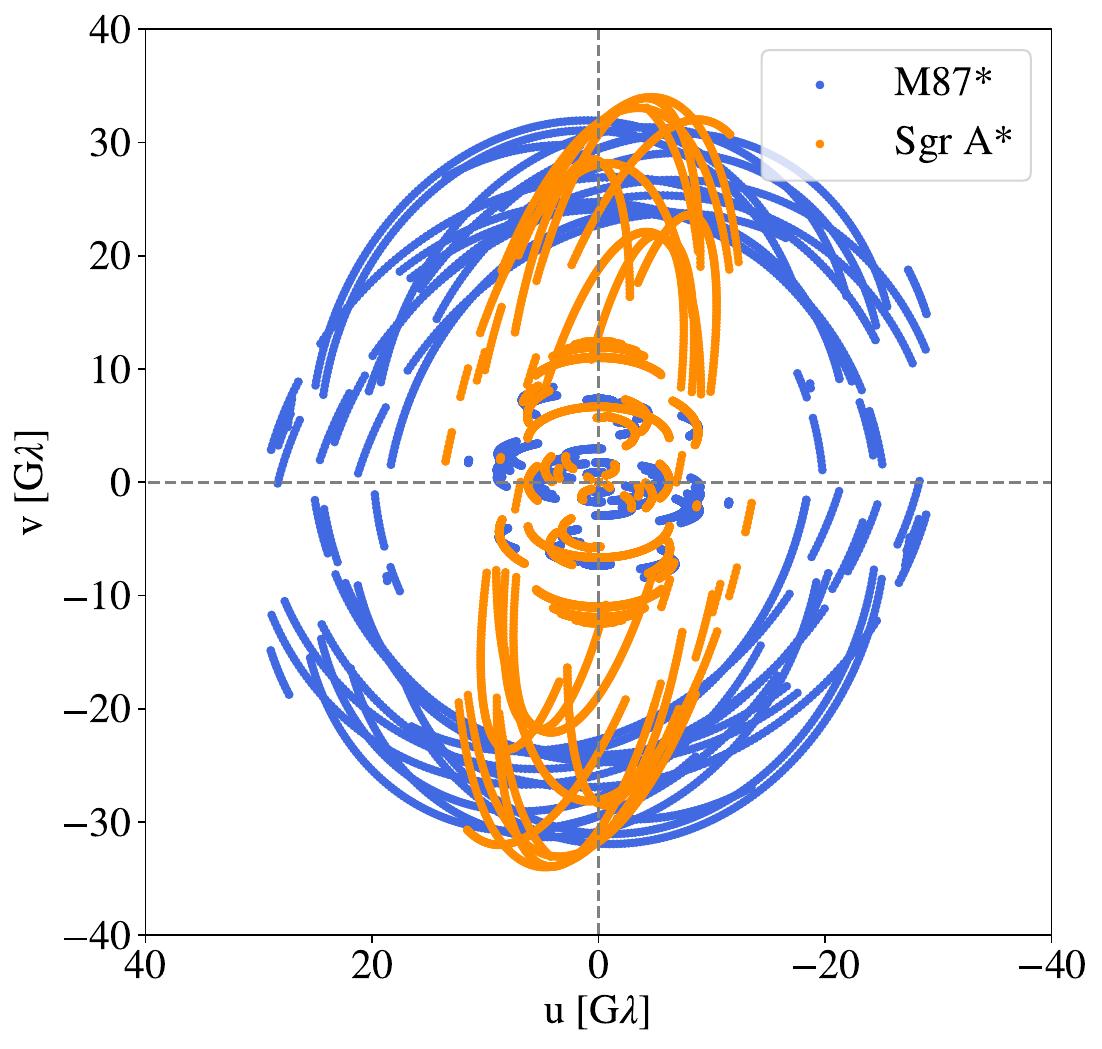}
    \caption{24~hour \emph{(u,v)} coverage of M87* and Sgr\,A* by \BHEX on 1\textsuperscript{st} Jan and 1\textsuperscript{st} June, respectively. Observations limited to when an optical downlink with the ground is achievable. Terminal gimbal capability limited to \(\pm\)70\(\degree\). 36.0\% and 23.4\% loss of coverage for M87* and Sgr\,A*, respectively.}
    \label{f:CDHuv}   
\end{figure}

The proposed configuration does require that the optical terminal is mounted such that the radio telescope antenna surface is not within its FOV. The terminal would need to be mounted such that it protrudes past the antenna surface. A more detailed mechanical analysis is required to assess the feasibility of this. Furthermore, if the gimbal capability is any less than the optimal \(\pm\)90\(\degree\), the positions in the orbit at which a downlink with the ground can be achieved begin to decrease. with a gimbal capability of \(\pm\)70\(\degree\), the \emph{(u,v)} coverage of M87* and Sgr\,A* is reduced to that depicted in Fig. \ref{f:CDHuv}.

Even if the data downlink constraint is mitigated through one of the methods discussed previously, there is still the challenge of tracking for fine orbit determination. Tracking of the spacecraft is required to accurately reconstruct the spacecraft's orbit so that its position is known throughout observations for the VLBI correlation process. A network of ground tracking stations is required to achieve this for mission orbits above altitudes where Global Navigation Satellite Systems (GNSS) can be used. For \emph{BHEX}, the preliminary intention is that tracking will be performed through a series of radio frequency ground stations, in addition to the optical network. However, depending on the coverage of the spacecraft this set of ground stations can achieve, fringe finding during correlation may not be possible for parts of the spacecraft's orbit with insufficient position accuracy. In application to \emph{THEZA}, this challenge is described in \cite{gurvits_science_2022}.

If an onboard frequency standard is not included in the spacecraft design, a link with the ground is required throughout observations to provide a stable reference signal, as was the case with VSOP-HALCA \cite{murata_vsophalca_2000}. This constraint is removed if a frequency standard is flown onboard, as demonstrated by RadioAstron with a Hydrogen Maser \cite{kardashev_radioastron-telescope_2013}. The current baseline scenario for \BHEX involves an onboard crystal ultra-stable oscillator \cite{marrone_black_2024}.

\subsection{Thermal Constraints}
\label{ss:thermal}
\noindent For space-based VLBI it is essential that the Sun does not illuminate the radio telescope antenna opening as this can result in heating of the critical components in the radio signal chain, which must be thermally controlled through means of a Sun shield and/or  active thermal control. The Sun also produces radio signals and therefore is a source of noise in the science measurements.

As such, it is required that the Sun does not fall within the cone of avoidance of the antenna when observations are being performed. For example, RadioAstron could not perform an observation if the Sun fell within 90\(\degree\) of the antenna boresight \cite{radioastron_science_and_technical_operations_group_radioastron_2019}. For the same reason, there are typically limitations on the Earth and sometimes Moon position during observations, due to its thermal emission. RadioAstron could not observe if the Earth limb or Moon were within 5\(\degree\) of the antenna boresight \cite{radioastron_science_and_technical_operations_group_radioastron_2019}.

Various strategies can be used to reduce this Sun exclusion requirement such as the use of Sun shields and shading of the antenna surface. However, these approaches can add significant mass and complexity to a mission, particularly in the case of a Sun shield which will often require a deployable mechanism, such as that used for the James Webb Space Telescope (JWST) \cite{rigby_science_2012}. For analysing the impact of the Sun exclusion angle on \BHEX and \THEZA observation operations, a 90\(\degree\) limitation between the solar direction and antenna pointing is considered as this is not dependent on additional Sun shields or antenna shading systems. 5\(\degree\) Earth limb and Moon exclusion angles are used as preliminary values for \BHEX analysis.

As would be expected, for half of the year the angle between the antenna boresight and the Sun is greater than 90\(\degree\) for either primary target source (see Fig. \ref{f:solarAngle} in section \ref{ss:power} for illustration of this). Therefore, observation of any source is restricted to a 6~month period each year, without consideration of any other limitations. For the \BHEX mission, this observing \emph{season} is further restricted to times of the year which are most favourable for observations from the ground sites' perspective. \BHEX is planning to observe M87* and Sgr\,A* for 3~month periods between January-April and June-September, respectively \cite{BHEX-2024-SPIE}. These are therefore the times of year for which other elements of the spacecraft design should be optimised for, to minimise the impact of the functional constraints on observations.

As is evident from the attitude sphere depicted in Fig. \ref{f:CDHsphere}, for part of the orbit, the Earth falls within 5\(\degree\) of the antenna boresight when observing Sgr\,A*. This occurs when the source is blocked by the Earth and therefore observations cannot take place anyway. However, during this period, for around 3.75~hours every orbit, the radio chain will be heated up by the Earth's thermal flux if the spacecraft is kept pointed in the Sgr\,A* direction. A full thermal analysis will be necessary to calculate the duration required for a cooling period before observations can commence after a period of time spent behind the Earth.

To increase the sensitivity of the instrument, the receiver system on \BHEX will be kept at \(\sim\)4.5~K during observations via cryogenic cooling \cite{rana_black_2024}. The cryocooler will require two radiator surfaces for operation: a deep-space pointing, passive heat rejection stage and a warm end heat rejection surface. The latter could occasionally be Sun/Earth pointing, the limits of which will depend on a more detailed thermal analysis of the mission. The heat rejection stage must be deep-space pointing throughout observations. The strategy proposed in section \ref{ss:CDH}, whereby the spacecraft is rotated every half orbit period to keep one side approximately Earth-facing, is highly beneficial for the thermal control of the system. Assuming that the optical terminal is pointed along the +X axis, positioning of the radiator surface on one of the perpendicular body-axis could provide the required deep-space pointing for the majority of the time. 

However, even this configuration would result in occasional appearance of the Moon in the radiator's FOV. At lunar perigee, the angular size of the Moon from \emph{BHEX}'s orbit is only 0.55\(\degree\). Assuming a solar irradiance of 1361 $W/m^2$ and a Moon albedo coefficient of 0.12, the thermal output of the Moon in \emph{BHEX}'s orbit is only 0.004 $W/m^2$. A more detailed thermal analysis would be required to confirm whether the Moon could be within the radiator's FOV and the required thermal control still be achieved. For this analysis, it is assumed that the Moon's thermal output is low enough such that it can be within the FOV of the radiator.

The ideal pointing for a radiator surface would be in the antenna direction as this will always be pointed towards deep-space during observations. However, with a 3.5~m antenna diameter, this would very likely require a deployable structure to provide an unobstructed deep-space view for the radiator. It is highly preferable to avoid such mechanisms, as they add mass, complexity and cost to the mission. With this constraint in mind, there is no available position for a radiator surface for which the Earth or Sun doesn't fall within 90\(\degree\) of its normal vector, when observing M87* or Sgr\,A*. To minimise the impact of blinding of the radiator surface on observations, some baffling of the radiator will be required to shield it from the Earth and Sun.

\begin{figure}[t!]
    \centering
    \includegraphics[width=\columnwidth]{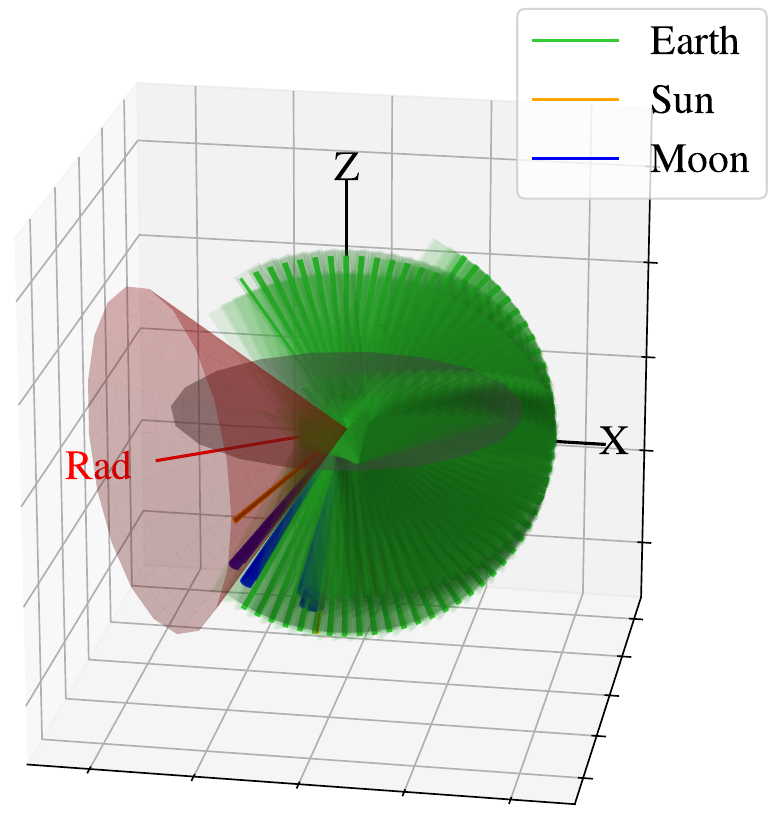}
    \caption{Attitude sphere showing optimal \BHEX radiator configuration if a baffle with a \(\pm\)40\(\degree\) FOV is implemented to shade the surface. Sun, Earth and Moon directions during observation of M87* and Sgr\,A* (1\textsuperscript{st} Jan and 1\textsuperscript{st} June, respectively.) are also shown.}
    \label{f:RadSphere}   
\end{figure}

Fig. \ref{f:RadSphere} depicts the optimal position for a radiator surface with a baffle limiting its FOV to \(\pm\)40\(\degree\), calculated using the process in section \ref{s:optimisation}. The attitude sphere shows the Sun, Earth and Moon positions for only a single day of observations for clarity. In reality, the optimal radiator positioning has been determined considering the wider range of Sun and Moon positions across the full observation season of each source. As can be seen in Fig. \ref{f:AttitudeSphere}, the Sun and Moon positions form large arcs across the attitude sphere during the observation season, further limiting the acceptable positions for a radiator.

\begin{table}
\small
\caption{Loss of \emph{(u,v)} coverage during observation seasons of M87* and Sgr\,A* (i.e. the time that the radiator is not deep-space pointing). Variation shown for different radiator baffle angles, defined as a cone around the radiator surface normal vector. Use of a different baffle angle results in the optimal location for the radiator varying.}
\label{t:radBaffle}       % Give a unique label
\begin{tabular}{c|c|c|c}
\hline\noalign{\smallskip}
Baffle Angle [\(\pm\)\(\degree\)] & Radiator Normal & M87* & Sgr\,A* \\
\hline
40 & [-0.809, -0.588, 0] & 0.3\% & 6.7\% \\
50 & [-0.809, -0.588, 0] & 38.6\% & 10.3\% \\
60 & [0, 1, 0] & 13.6\% & 65.5\% \\
70 & [0.309, 0.951, 0] & 17.9\% & 74.0\% \\
80 & [0, 1, 0] & 23.9\% & 90.96\% \\
90 & [-0.809, 0.588, 0] & 67.6\% & 89.8\% \\
\hline
\end{tabular}
\end{table}

The radiator configuration shown in Fig. \ref{f:RadSphere} would be deep-space pointed for the majority of the M87* and Sgr\,A* observation seasons. Observation of M87* is unaffected but for \(\sim\)6~days of the Sgr\,A* season, the radiator would be blinded by the Sun. However, increasing the baffle angle results in the radiator being pointed towards the Earth or Sun for longer periods of time, during which observations cannot be performed. The optimal location of the radiator surface also varies if less shading is implemented. Table \ref{t:radBaffle} shows the percentage loss in \emph{(u,v)} coverage of M87* and Sgr\,A* during their observation seasons due to the radiator not being deep-space pointed. How this varies with baffle angle is also provided along with the optimal radiator position for a given baffle angle. \(\pm\)90\(\degree\) is a radiator surface with no baffle / shading. Some of the proposed radiator positions are only suitable if the antenna surface obscuring its FOV is not impactful or, if it is, the radiator can be mounted in such a way as to remove the obstruction.

Positioning of the radiator surfaces on a space-based VLBI mission is a challenging issue and one that cannot simply be solved by adding more radiators without incurring significant mass and cost increases. Baffling of the radiator requiring deep-space pointing is likely to be essential for minimising the impact on observations to an acceptable level. Further thermal analysis is required to consider the positioning of the warm end rejection surface to determine what level of Sun exposure is acceptable.

\subsection{Attitude and Orbit Control}
\label{ss:AOCS}
\noindent The Attitude and Orbit Control System (AOCS) of a space-based VLBI mission places a number of functional constraints on when observations can be performed. Space VLBI requires highly accurate and stable attitude control in order to point a very narrow antenna beam towards the target source. As such, the attitude control accuracy needs to be on the order of arcseconds (see RadioAstron attitude control: \(\pm\)10" \cite{radioastron_science_and_technical_operations_group_radioastron_2019}). This requires utilisation of a stellar-gyro system using star trackers and gyroscopes (the most accurate attitude determination method available), and reaction wheels for attitude control. This is a standard attitude control method for space astronomy missions with similar pointing requirements (see James Webb Space Telescope, Hubble, etc. \cite{rigby_science_2012}).

A star tracker is a digital camera that maps the stars observed within its FOV to an internal star catalogue. By identifying the stars in the image, a star tracker can estimate the spacecraft attitude by determining the orientation of the star field with respect to the Earth Centered Inertial (ECI) frame \cite{Fundamental_AOCS}. Star trackers exhibit the highest error in orientation estimation about their boresight (i.e. pointing direction). In order to achieve arcsecond-level estimation at a system level, this typically requires two star trackers, mounted at at least 45\(\degree\) to each other \cite{radioastron_science_and_technical_operations_group_radioastron_2019,Fundamental_AOCS}. Star trackers can be blinded when the Sun, Earth and sometimes Moon, fall within their FOV. At such times, they cannot provide an attitude estimation and therefore for space VLBI, observation cannot take place.

\begin{figure}
    \centering
    \includegraphics[width=\columnwidth]{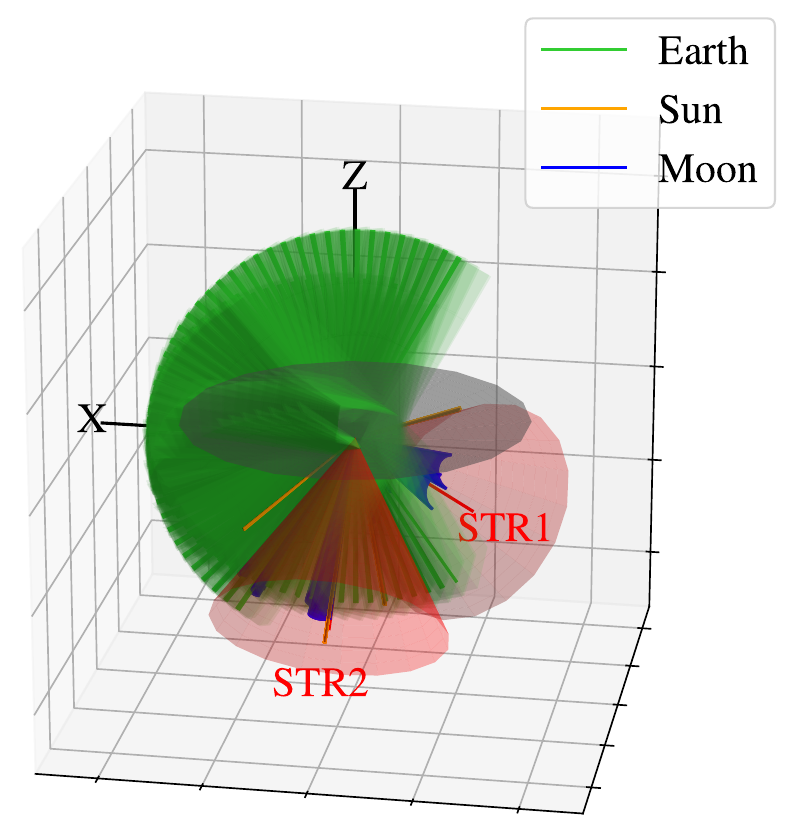}
    \caption{Attitude sphere showing optimal \BHEX star tracker (STR) configuration and Sun, Earth and Moon positions during observation of M87* and Sgr\,A* (1\textsuperscript{st} Jan and 1\textsuperscript{st} June, respectively).}
    \label{f:STRsphere}   
\end{figure}

Consider the \BHEX mission, utilising a minimum of two star trackers to provide the required attitude determination. The star trackers must be positioned to minimise the times during a given observation season that they are blinded by the Sun, Earth or sometimes Moon (many modern star trackers can operate with the Moon in the FOV). The optimisation problem is complicated by the fact that the star tracker blinded times should be minimised for both M87* and Sgr\,A*. If the operational approach described in section \ref{ss:CDH} is utilised, the spacecraft is rotated by 180\(\degree\) each half orbit period to keep the Earth within the optical terminal FOV. The real-time downlink requirement is the most constraining element of \emph{BHEX}'s design so it is likely that minimising its impact will drive the concept of operations. Using this method, the Earth is kept within one half of the spacecraft's attitude sphere, as shown in Fig. \ref{f:STRsphere}. This is in fact beneficial for the placement of components that require deep-space pointing as one side of the spacecraft would always be kept away from the Earth.

\begin{figure}[t!]
    \centering
    \includegraphics[width=\columnwidth]{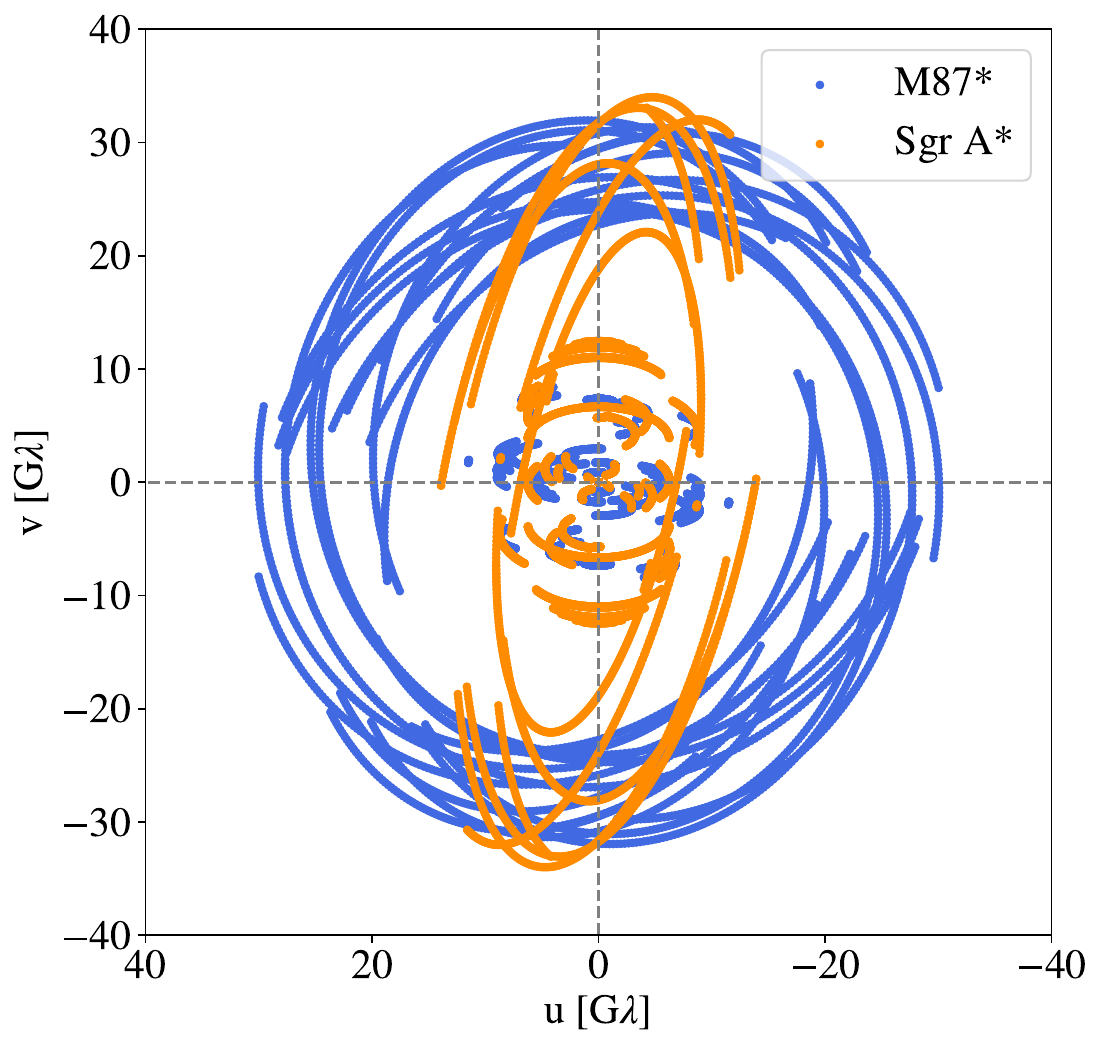}
    \caption{24~hour \emph{(u,v)} coverage of M87* and Sgr\,A* by \BHEX on 1\textsuperscript{st} Jan and 1\textsuperscript{st} June, respectively. Observations limited to times when neither star tracker is blinded. 17.1\% and 51.7\% loss of coverage for M87* and Sgr\,A*, respectively.}
    \label{f:STRuv}   
\end{figure}

\begin{figure*}
    \centering
    \includegraphics[width=0.8\textwidth]{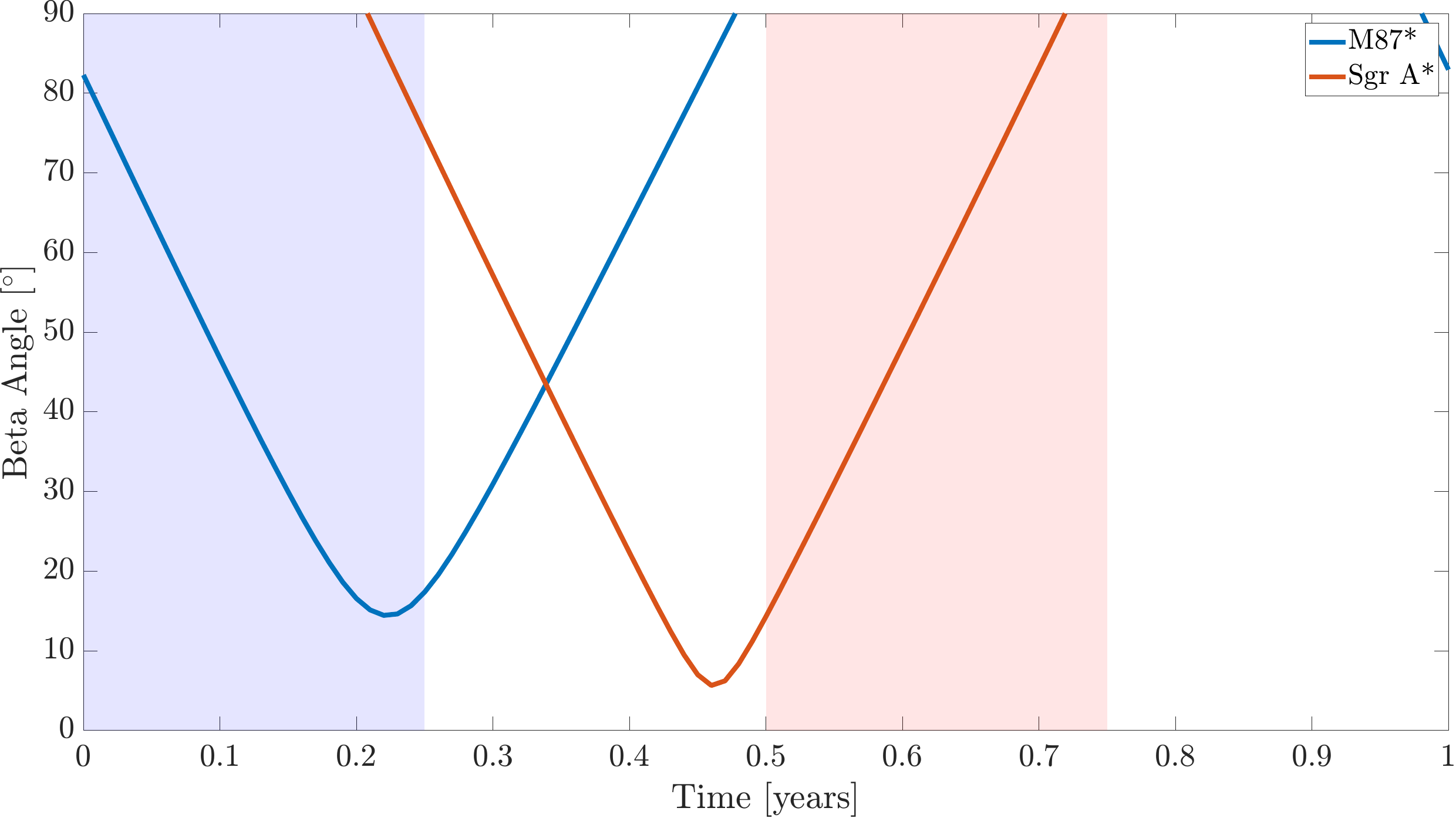}
    \caption{Angle from solar panel normal pointed in the negative Z-axis to the incident solar radiation (beta angle). Variation shown for observation of M87* and Sgr\,A*. Observation seasons for both sources highlighted.}
    \label{f:solarAngle}   
\end{figure*}

However, as the spacecraft is rotated 180\(\degree\) about the Z-axis each half orbit, the Sun positions on the attitude sphere are effectively doubled. This can be seen in Fig.~\ref{f:STRsphere} as there are four distinct Sun position arcs for observing M87* and Sgr\,A*, rather than the expected two. This complicates the star tracker positioning. With the antenna  pointed in the $+$Z axis, unless the star trackers are mounted such that they are not obscured by the antenna surface, it is reasonable to assume that they cannot be mounted with the boresight pointing in a $+$Z direction. This further constrains the available position of the star trackers to the $-$Z hemisphere of the attitude sphere. The attitude sphere in Fig.~\ref{f:STRsphere} depicts the optimal locations for two star trackers, under the array of conditions stated above. These positions offer minimal blinding during the observation season of each of the primary sources, whilst meeting the criteria of having a 45--90\(\degree\) separation between their boresights. Sun blinding is avoided completely as this would result in entire days during which observations would not be possible. The unit vectors of the star trackers in the body-fixed frame are: [-0.476, -0.655, -0.589] and [0, 0.707, -0.707]. Fig. \ref{f:STRuv} shows the \emph{(u,v)} coverage of M87* and Sgr\,A*, when star tracker blinding constrains observation times.

Mitigation of the star tracker functional constraint could be achieved by introducing a third star tracker in a hot redundant state. This unit could then be positioned to provide coverage for when either of the other star trackers are blinded. Such a decision illustrates why analysis of the functional constraint impact is crucial early in the spacecraft design process. However, this configuration optimisation has considered only M87* and Sgr\,A*. Observation of other sources adds another set of Earth, Sun and Moon positions to the attitude sphere which would make the positioning of the star trackers even more complex (see section \ref{s:scienceReturn} for a more detailed discussion on this). 
Other functional constraints related to the AOCS exist but are not discussed in detail in this section. These constraints will place additional limitations on when observations can be performed, although they are unlikely to be as stringent as the star tracker placement.
\begin{itemize}[noitemsep]
\item Breaks in observations during reaction wheel desaturation to dump accumulated angular momentum
\item Performance of orbit station-keeping or transfers, during which a fixed antenna cannot be pointed at the target source
\end{itemize}

\noindent In relation to concepts such as \emph{THEZA}, the impact of the star tracker constraint can be massively reduced if the mission is operated far from the Earth. As shown clearly in Fig. \ref{f:STRsphere}, the Earth covers the greatest area of the attitude sphere for a mission such as \BHEX.

\subsection{On-board Power Subsystem}
\label{ss:power}
\noindent The payload of a space-based VLBI mission, consisting of: multiple receivers, cryogenic cooling system, ultra-stable oscillator and an optical communications terminal, will have a considerable power requirement. The optical communications system on TBIRD alone (which the \BHEX terminal will be based on), has a power requirement in excess of 100~W \cite{schieler_-orbit_2023}. Whilst the power system design will be very dependent on the specific mission, a high-level investigation into the geometry of the Sun and the likely solar panel configuration of such a spacecraft can be performed to investigate the functional constraint the power requirement may place on the mission.

As described in section \ref{ss:thermal}, observation of a given source requires that the angle between the main antenna pointing and the Sun direction is greater than 90\(\degree\) (perhaps slightly less if some shading of the antenna is included). Therefore, the logical position for the solar panels is in the opposite direction to the antenna as this is approximately where the Sun will be located during observations. As such, consider a fixed solar panel design pointing in the negative Z-axis. The power generated by a solar panel is dependent on the characteristics of the solar cells utilised, the panel area and incidence angle of the Sun. The power generated varies in proportion to the cosine of the incidence angle of the solar radiation, measured from the panel normal. Fig.~\ref{f:solarAngle} shows the variation in this angle throughout the year, when observing either of the main sources.

Solar panels suffer from performance degradation throughout their lifetime due to damage from radiation and micrometeorite impacts. Therefore, they must be sized such that they provide sufficient power at the end of the spacecraft's lifetime, considering these degradation factors. In order to provide the required power to a VLBI payload when the solar incidence angle is not 0\(\degree\) with respect to the panel normal, the solar panel area will need to increase by a factor equal to the reciprocal of the cosine of the incidence angle. 

For \emph{BHEX}, observations of M87* and Sgr\,A* are planned to take place during specific seasons, during which the Sun avoidance angle requirement is met. A similar strategy is likely to apply to any space-based VLBI mission observing these sources. As can be seen in Fig. \ref{f:solarAngle}, for \(\sim\)30\% of the M87* observation season, the solar incidence angle on a $-$Z panel is greater than 60\(\degree\). For this period of time, the power generation of this panel would be 50\% of that when the solar incidence angle is 0\(\degree\) and thus, the panel would need to be twice the area for observations to take place.

As such, there is a clear trade-off between the configuration of the solar panels and the times of year at which sufficient power is provided to the VLBI payload in order to conduct an observation. Several mitigation strategies exist for this constraint:
\begin{itemize}[noitemsep]
    \item{Additional, fixed solar panels, pointing in different directions}
    \item{Use of steerable solar panel(s)}
    \item{Shift the observing season such that the angle on a $-$Z facing panel is closer to zero for a longer period}
    \item{Operational mitigation: periodic interruptions in observations to point the solar panels towards the Sun and charge up batteries}
\end{itemize}

\noindent Each of these strategies require the resolution of a trade-off with other elements of the system and its operation. The addition of solar panels or the use of steerable systems adds significant mass to the system. Steerable panels also introduce considerable complexity as the mechanism is a mission-critical component. Furthermore, they may be outside of the envelope of a NASA SMEX mission for \emph{BHEX}. The observation season is also driven by factors beyond just this power constraint (see section \ref{ss:thermal}).

The power constraint will need to be overcome for any space-based VLBI mission. Although this is indeed a design challenge that applies to all space observatories, it will likely still limit when observations can be performed.

\section{Mitigating Impact on Science Return}
\label{s:scienceReturn}
\begin{figure}
    \centering
    \includegraphics[width=\columnwidth]{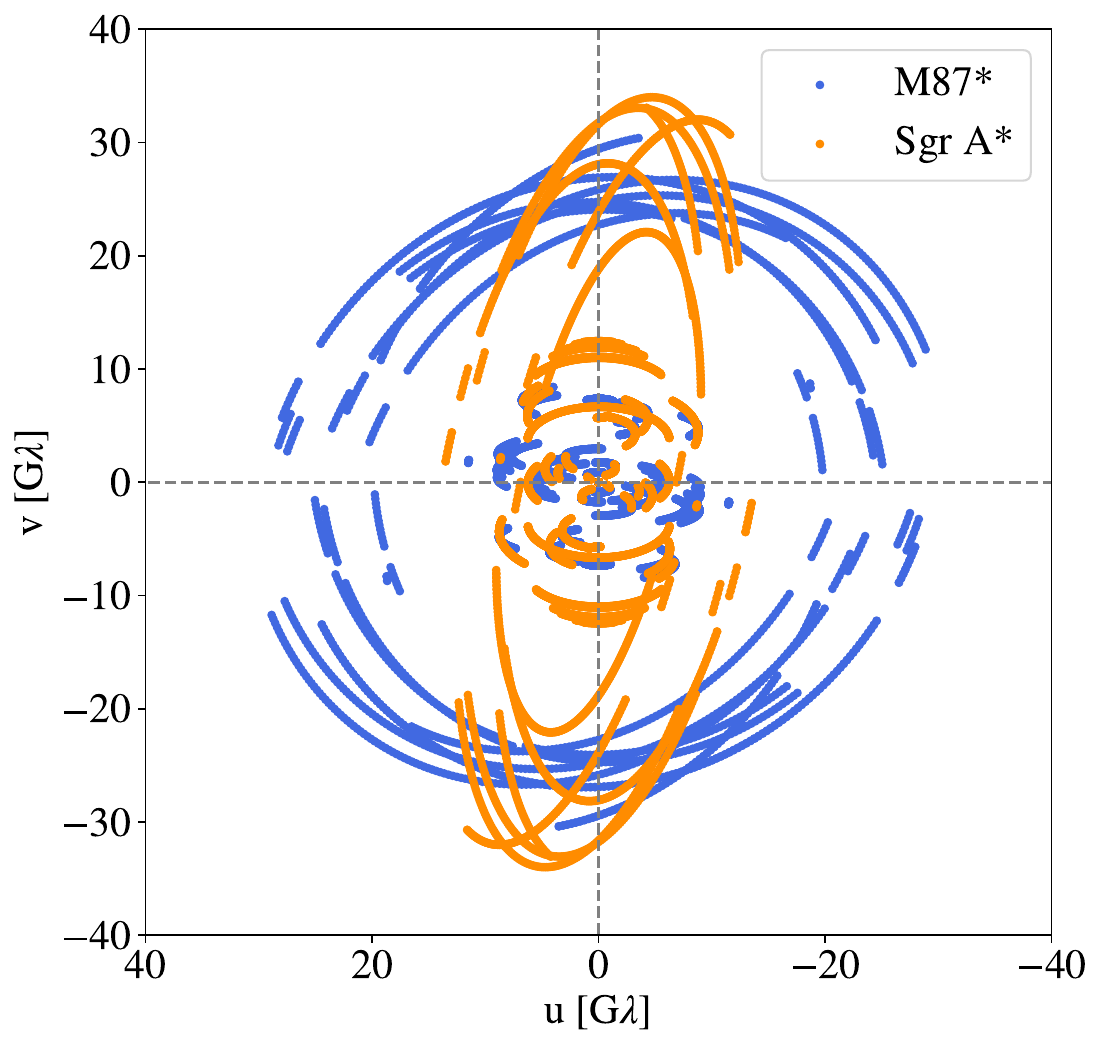}
    \caption{24~hour \emph{(u,v)} coverage of M87* and Sgr\,A* by \BHEX on 1\textsuperscript{st} Jan and 1\textsuperscript{st} June, respectively. Observations limited to times when neither star tracker is blinded, the radiator surface is deep-space pointed and a real-time downlink with the ground can be maintained (\(\pm\)70\(\degree\) gimbal capability). 63.9\% and 61.8\% loss of coverage for M87* and Sgr\,A*, respectively.}
    \label{f:CombUV}   
\end{figure}

\noindent So far, each of the presented constraints have been analysed in isolation. The \emph{(u,v)} coverage figures showing the impact on observations have not included the effect of the other constraints. In reality, the functional constraints will stack up, resulting in a boolean product of the time intervals at which observations can be performed. It is the combination of the constraints that will drive the availability of the system and thus, the science return of the mission. Fig. \ref{f:CombUV} depicts the \emph{(u,v)} coverage achieved by \BHEX when the three primary constraints described in this paper (real-time downlink, radiator surface pointing, star tracker blinding) are all impacting observations at the same time. As can be seen, the total loss in \emph{(u,v)} coverage is far more significant than the individual impact of each constraint.

The future space-based VLBI missions must of course be designed to minimise the impact of these (and other) functional constraints. However, as has been shown in the previous sections, the complexity of performing VLBI in space means that the number of functional constraints makes it highly unlikely that they will all be completely mitigated. As has been discussed, strategies exist to reduce, and in some cases, negate the impact of these constraints. For example, inclusion of additional star trackers, positioned in such a way as to have at least two that are never blinded. Various methods have been proposed for removing the real-time optical downlink constraint; mass data storage being the obvious solution. The difficulty is in implementing these strategies within the mass, power and and financial envelope of the mission. A typical mass limit of a NASA SMEX spacecraft is $\sim$200-300~kg \cite{nasa_smex}. This, together with other programmatic constraints of a SMEX mission are likely to be the limiting factors in overcoming the functional constraints of the mission. 

Furthermore, the constraint analysis in this investigation has focused on the two largest sources as seen from the Earth: M87* and Sgr\,A*. As described in section \ref{ss:Science}, the science applications of VLBI require wider observations than these two SMBH targets. It has been shown how the attitude sphere can provide a clear depiction of the constraint optimisation problem. Fig.~\ref{f:CDHsphere}, \ref{f:RadSphere} and \ref{f:STRsphere} show the geometry of the Earth, Sun and Moon throughout observation of M87* and Sgr\,A*. Observation of each additional source adds another set of Earth, Sun and Moon positions to the attitude sphere that must be taken into consideration when optimising the spacecraft configuration.

\begin{figure*}[t]
    \centering
    \includegraphics[width=\textwidth]{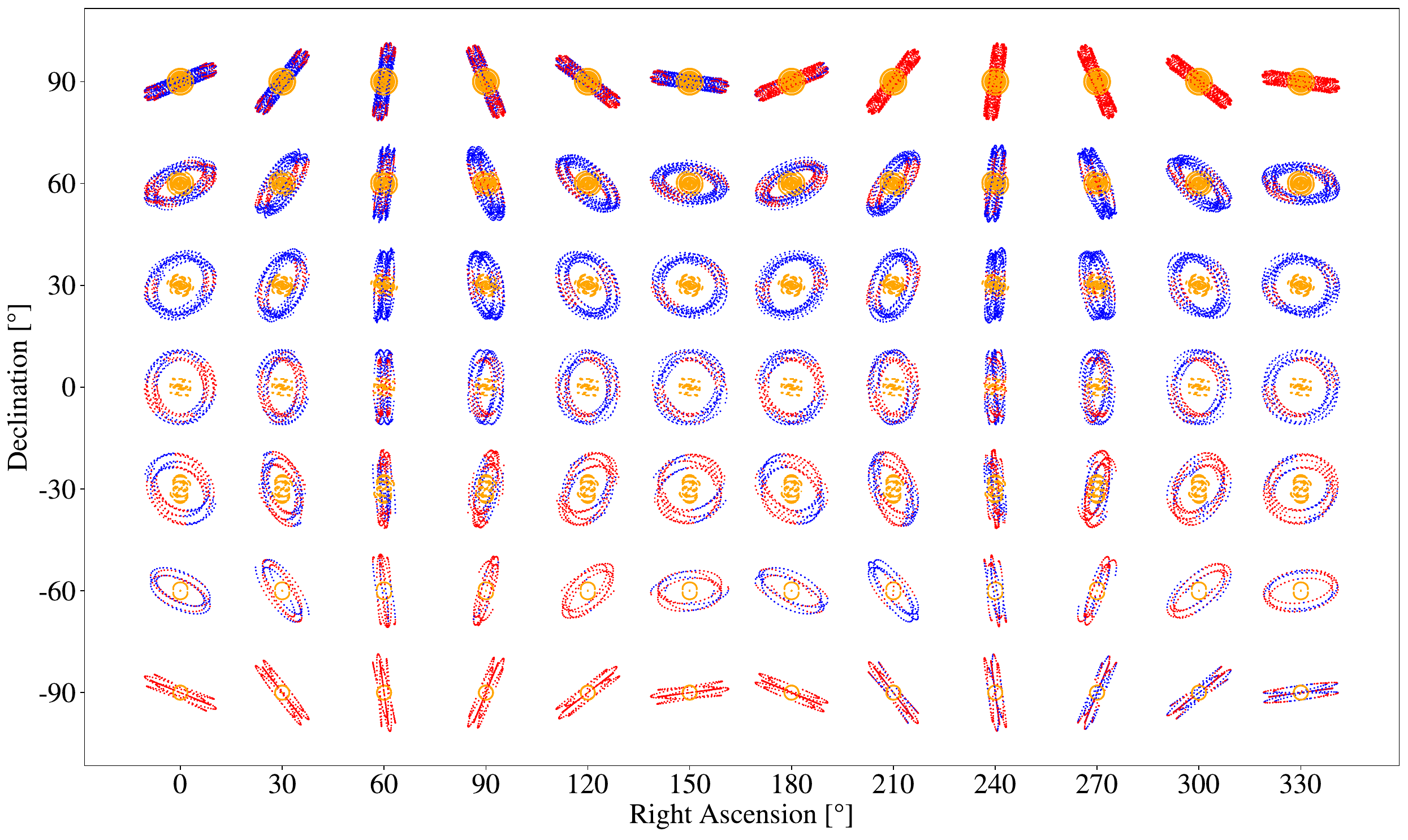}
    \caption{\emph{All-sky (u,v)} coverage achieved by \BHEX of targets across the celestial sphere. The star tracker and radiator functional constraints have been included in this calculation, with the units positioned in the `optimal' locations determined in sections \ref{ss:thermal} and \ref{ss:AOCS}. Observation of each right ascension, declination position is conducted at the time of year when the source is 180\(\degree\) separated from the Sun.}
    \label{f:allsky}   
\end{figure*}

Fig.~\ref{f:allsky} depicts an \emph{all-sky} \emph{(u,v)} plot for \BHEX which shows the coverage achieved by the full interferometer of targets across the celestial sphere. This coverage has been calculated with inclusion of the functional constraints imposed by two star trackers and a radiator surface. Coverage lost due to the functional constraints limiting observations is shown in red. The figure demonstrates how configurations that are `optimal' for observing M87* and Sgr\,A*, are insufficient for other sources, resulting in very significant losses in \emph{(u,v)} coverage. The figure also includes the effect of a 90\(\degree\) Sun exclusion and 5\(\degree\) Earth limb and Moon exclusion, the former resulting in total loss of coverage for some targets at high declinations.

Under the exclusion angle assumptions made in the previous sections, positioning of two star trackers and a radiator surface so that they are not blinded throughout observations is impossible given even one additional source to M87* and Sgr\,A*. Even the strategy described in section \ref{ss:AOCS}, flying a third star tracker, will not be sufficient as other target sources of \BHEX are added to the attitude sphere. For the real-time downlink constraint, it has been shown that the impact on \emph{(u,v)} coverage is highly dependent on the gimbal capability of the optical terminal. The effect of the real-time downlink constraint was not included in Fig. \ref{f:allsky} as to observe different sources, unique implementations of the attitude control strategy described in section \ref{ss:CDH} are required. For the generation of Fig. \ref{f:allsky}, the optimal attitude control strategy to maintain the real-time downlink when observing M87* was implemented. Again, this emphasises that what is optimal for M87* and Sgr\,A* does not guarantee that the same observing strategies can be used by \BHEX for other sources.

As such, for \BHEX and future space-based VLBI missions, additional methodologies will be required to overcome the complex optimisation problem that is constraint mitigation. The spacecraft design and mission architecture must be optimised for observation of a set of primary sources, for which the \emph{(u,v)} coverage will be maximised. Sources located close together in the sky will be subjected to very similar constraints, as the Earth, Sun and Moon profiles on the attitude sphere will be similar. For example, 3C273 and 3C279, quasars of high scientific value for studies with \emph{BHEX}, lie close to M87*. Therefore, optimising the mission for observation of carefully selected regions of sky will increase the number of sources for which the impact of constraints can be mitigated.

As described previously, another promising strategy for constraint mitigation lies in the attitude control of the spacecraft. The elements of the design for which functional constraints exist are usually based around a specific relationship with the Earth, Sun or Moon. For some constraints, this relationship is positive: the optimal location of an optical terminal is the point on the attitude sphere which provides the greatest intersection with the Earth sphere. For others it is negative: star trackers and radiator surfaces blinded by Earth, Sun (and Moon). This polarity can be taken advantage of by varying the attitude of the spacecraft throughout observations, as was described as the primary mitigation strategy in section \ref{ss:CDH}. Proposed was a 180\(\degree\) rotation about the Z-axis, every half orbit period. However, if the gimbal capability of the terminal is less than the optimal \(\pm\)90\(\degree\), rotating the spacecraft constantly throughout an observation to keep one side near-Earth facing would increase the times at which a link with a ground station can be achieved. In contrast, as was shown in sections \ref{ss:thermal} and \ref{ss:AOCS}, star trackers and radiator surfaces would be provided with a mounting surface which will never be pointed towards the Earth.

The difficulty in this operation lies in the accurate pointing required of the antenna towards the target source. If rotation is conducted continuously, the spacecraft centre of mass would have to be stringently positioned to ensure the rotation axis is exactly about the antenna pointing. Furthermore, maintaining the required pointing accuracy and stability whilst rotating would be a significant challenge. A perhaps more realistic approach is to conduct discrete rotations at set intervals around the orbit, settling into an inertially-fixed attitude to perform observations. This is more feasible from an attitude control perspective but observation time would be lost during the slew and subsequent settling of the spacecraft attitude control. Variation in the polarisation of the received signals would also have to be corrected for on the ground with estimations of the spacecraft's attitude. With a control accuracy of \(\sim\)arcseconds, the attitude knowledge could be expected to be an order of magnitude higher than this. This strategy may be too ambitious for missions such as \emph{BHEX} with the constraints of a SMEX mission but it could be a legitimate option for future space VLBI missions.

Mission optimisation to minimise the impact of the functional constraints must also consider the required properties of the \emph{(u,v)} coverage that is achieved for each source, depending on the science objectives. The major benefit of space VLBI is the improvement in angular resolution due to the longer baselines and prospective ability to observe at wavelengths unachievable on the ground. To achieve this, the functional constraints cannot be allowed to impact observations on the longest baselines. It is also important that baselines with key ground stations, providing the highest SNR, are maintained. Optimisation of the mission to minimise the impact of constraints is therefore not only a spacial problem, but also temporal as the spacecraft must be designed so that unavoidable constraints occur at the least impactful times.

Furthermore, the total duration of \BHEX scans of a target source is expected to be \(\sim\)10~hours each day, considering all constraints related to planning observations. This may simplify constraint mitigation somewhat as there are less times during the day that the functional constraints must be minimised for. Observing for only \(\sim\)10~hours each day may well be a significant simplification for the power generation constraint as there would be a considerable amount of time between scans to recharge batteries for the next observation. This of course assumes that the batteries are sufficiently sized to provide the required power / power deficit for the duration of a single scan. However, if observations are to produce a \emph{(u,v)} coverage that provides  360\(\degree\) coverage of the \emph{(u,v)} plane, the same geometrical challenge in maintaining real-time downlink and avoiding star tracker / radiator blinding remains. 

\vskip 1em
\section{Conclusions and Outlook for Future Work}
\label{s:concl}
 
\noindent In this paper, we have demonstrated how the functional constraints pertaining to a space-based VLBI mission have a non-trivial impact on the achievable \emph{(u,v)} coverage and thus the potential science return. In doing so, we have shown that it is essential to consider the effects of these constraints as early in the mission concept development as possible. The \texttt{spacevlbi} Python package has been developed to enable analysis of such constraints and optimise the mission design to minimise their impact.

Space-based VLBI has the potential to advance large fields of astrophysics and fundamental physics which have until now been unattainable due to the limitations of ground VLBI. \BHEX will resolve the photon rings of M87* and Sgr\,A*, enabling precise measurements of mass and spin of these SMBHs. By achieving the finest angular resolution in the history of astronomy, \BHEX will provide invaluable contributions to other areas of research such as physics of inner areas of AGN and multi-messenger astronomy through spatially-resolving generators of gravitational wave emission (SMBHBs) \cite{BHEX-2024-SPIE}.

\BHEX is the most likely space VLBI concept to be realised in the near-future. However, as described in section \ref{ss:concepts}, numerous other mission concepts have been proposed. \THEZA in particular would provide an order of magnitude improvement in angular resolution compared to ground-based systems, forming space-space baselines with no theoretical limit on observing up to terahertz frequencies. A space-based VLBI mission is unavoidable if the science objectives discussed in this paper are to be met.

Although analysis of the functional constraints requires assumptions to be made about the spacecraft design and mission architecture, they should be considered as early in the mission concept development as possible to identify the areas which require the greatest attention and technological development. As the \BHEX concept matures, the process presented in this paper must be regularly repeated to determine whether functional constraints have been introduced/changed through design choices.

In this paper, optimisation of the system parameters to minimise the impact of the functional constraints has been primarily focused on the spacecraft configuration. It has been assumed that the orbit selection is almost completely driven by the science case, whereas in reality there will be other factors that contribute to the orbit design. In future work, optimisation of the functional constraint impact will include consideration of the spacecraft's orbit. The optimisation process presented in section \ref{s:optimisation} will be expanded to include selection of the orbital elements to minimise the impact of certain constraints. This is a process that the future space VLBI missions will have to go through during detailed design and such an optimisation technique would be highly valuable during that stage of future programmes.

The functional constraint analysis performed here has not considered more dynamic limitations on when observations can be performed. These could include constraints on how long certain pieces of equipment can be used for, either due to insufficient power provision or thermal implications. Such effects should be identified and their impact modelled as the design of \BHEX and other space VLBI concepts progress.

Investigation into the functional constraints and discussion of the difficulties associated with overcoming them is not intended to question the feasibility of space-based VLBI (past missions already prove its viability). It is to ensure that the future missions are as effective as possible and that the science return of these systems is maximised. It is often the case in human history that overcoming the most difficult of challenges opens the door for the greatest accomplishments, and space-based VLBI is no different.

\vskip 1em
\section*{Acknowledgements}
\label{s:ackno}

\noindent The authors would like to thank Michael Johnson and the rest of the \BHEX community for their ongoing efforts to realise this exciting mission. The authors are also grateful to Don Boroson, who helped define the attitude control strategy for maintaining real-time downlink.

SI is supported by Hubble Fellowship grant HST-HF2-51482.001-A awarded by the Space Telescope Science Institute, which is operated by the Association of Universities for Research in Astronomy, Inc., for NASA, under contract NAS5-26555.
\vskip 1em

\appendix
\section{\texttt{spacevlbi} Optimisation}
\label{s:A1}

\noindent In this section is described the methodology implemented in \texttt{spacevlbi} to determine optimal locations for spacecraft equipment whose performance is based upon some relationship with the Sun, Earth or Moon. The optimisation is performed using a single or multiple \texttt{spacevlbi} simulation(s). This enables the optimisation to be conducted for varying simulation properties, for example, observing different sources. The simulation results are used to evaluate the \emph{fitness} of each potential component position onboard the spacecraft. All possible positions for components are evaluated with the generation of candidate unit vectors, in which the component may point, that cover the entire attitude sphere.

\subsection{Spacecraft Attitude Definition}

\noindent The attitude matrix $A$ of the spacecraft is calculated using the TRIAD algorithm \cite{Fundamental_AOCS}
\begin{gather}
    A = b_{1}r_{1}^{T} + (b_{1} \times b_{x})(r_{1} \times r_{x}) + b_{x}r_{x}^{T} \\
    r_{x} = r_{1} \times r_{2} \\
    b_{x} = b_{1} \times b_{2} \\
    r_{1} = \frac{r_{source} - r_{sc}}{||r_{source} - r_{sc}||}
\end{gather}

\noindent Where $b_{1}$ is the spacecraft body axis pointed towards the target source unit vector $r_{1}$ in the spacecraft inertial frame. $r_{source}$ is the target source vector in ECI and can be calculated from the source right ascension $\alpha$ and declination $\delta$. $r_{sc}$ is the spacecraft position vector in ECI. A constraint body axis $b_{2}$ is pointed along a unit vector $r_{2}$ in the spacecraft inertial frame, perpendicular to the source direction. The constraint vector $r_{2}$ is a free parameter defining the roll $\theta$ of the spacecraft about the target source direction. The roll angle is measured from the vector perpendicular to $r_{1}$ that aligns closest to the celestial north pole.
\begin{gather}
    r_{north} = -(r_{1} \times b_{1}) \times b_{1}
\end{gather}

\noindent This vector is rotated by $\theta$ about the target source direction $r_{1}$ to calculate the final constraint vector
\begin{multline}
    r_{2} = r_{north}\cos{(\theta)} + (r_{1} \times r_{north})\sin(\theta) + \\ (1-\cos(\theta)(r_{1} \cdot r_{north})r_{1}
\end{multline}

\subsection{Optimisation Methodology}

\noindent For the optimisation process, consider the case of calculating the optimal position of a star tracker which requires that the Sun or Earth are not within the FOV ($\beta$ is the half-angle of the FOV) in order to function. A set of trial unit vectors in the spacecraft body frame are generated to cover the full attitude sphere. Each of these trial component positions $b_{test}$ are evaluated with the process below. The unit vector in the body frame is first rotated into the inertial frame
\begin{gather}
    r_{test} = A^{-1}b_{test}
\end{gather}

\noindent We define the following angles in the spacecraft inertial frame: $\phi$ is the angle between $r_{test}$ and the sun vector, $\psi$ is the angle between $r_{test}$ and the Earth limb vector and $\gamma$ is the angle between $r_{test}$ and the Moon vector. The fitness function $\Pi$ is calculated as
\begin{gather}
    \Pi = \sum_{\alpha=0}^{2\pi} \sum_{\delta=-\pi/2}^{\pi/2} \sum_{t=0}^{t_{sim}} f(t, \alpha, \delta, b_{test})
\end{gather}
\[
f(t, \alpha, \delta, b_{test})=
\begin{cases}
     1, & \psi(t, \alpha, \delta, b_{test})-\beta\leq 0 \\
                    & \phi(t, \alpha, \delta, b_{test})-\beta\leq 0 \\
                    & \gamma(t, \alpha, \delta, b_{test})-\beta\leq 0 \\
    0, & \text{otherwise }
\end{cases}
\]

\noindent This function calculates the total number of time steps, during a simulation of duration $t_{sim}$, for which the Sun, Earth or Moon condition is violated for the component in question. The function also includes the effect of varying the target source position in right ascension and declination. The optimal solution is then the vector $b_{test}$ for which $\Pi$ is a minimum. The form of the fitness function subtly changes depending on the component being optimised. Table \ref{t:FitnessFunctions} describes qualitatively this variation.

\subsection{Optimisation Performance}
\noindent Various methodologies have been proposed for spacecraft configuration design optimisation (see \cite{silva_multiobjective_2019, Star_Tracker_Opt} for similar approaches). These methods optimise only specific component positions: radiators and star tracker, respectively. Furthermore, most optimisation processes utilise a genetic algorithm which mimics natural evolutionary processes on an initial population of potential solutions to converge on an optimal result. Genetic algorithms are an effective way to optimise multi-parameter, multi-objective problems with a large search space, but they can miss optimal solutions by finding local rather than global minima.

The optimisation approach presented here evaluates the full search space of possible component positions (at the expense of runtime, discussed further below), for a range of component types. Furthermore, the optimisation has been designed for astronomy missions which require pointing at various target sources across the celestial sphere, and not just a single attitude state.

The impact of the functional constraints is dependent on multiple, unrelated parameters (e.g. target source position, Moon/Sun position at a given date and time). Therefore, searching of the entire search space is unavoidable in order to find optimal configurations that are suitable for the wide ranges of these input parameters that will be experienced during the mission. The nature of this optimisation has the potential to result in long runtimes. For example, simulation of a single day of observation at a time step of 200~s takes \(\sim\)90~s of real time (when run with only J2/J3 orbit perturbations and the configuration depicted in Fig. \ref{f:bhexOrbit}). Extending this to simulation of a full year (to capture all possible Sun/Moon positions) would take \(\sim\)9~hours (executed on a 10 core laptop computer). This is only for observation of a single source.

To perform efficient optimisations that do not require impractical run times, the evaluation parameter space must be reduced by considering other factors. For example, execution across an entire year is unnecessary. Instead, the simulation can be performed for a single day at discrete times throughout the year to capture the full variation in Sun and Moon positions. The attitude sphere can then be used to ensure that interpolating the Sun and Moon positions does not cause a functional constraint with the calculated, optimal component configuration. Furthermore, if some of the target sources are located at similar positions in the sky, a single optimisation run may be sufficient for those sources.

\begin{table}
\small
\caption{Physical interpretation of fitness functions for optimising configuration of specific spacecraft components.}
\label{t:FitnessFunctions}       % Give a unique label
\begin{tabular}{m{1.5cm}|m{6cm}}
\hline\noalign{\smallskip}
Component & Fitness Function \\
\hline
Star \newline Tracker & Minimise time the Sun, Earth (or Moon) are within the FOV \\ \hline
Radiator & Minimise time the Sun, Earth or Moon can irradiate surface \\ \hline
Optical \newline Terminal &  For real-time downlink, maximise time that link can be achieved with the ground \\ \hline
Solar \newline Panel &  Maximise time that incident solar angle is less than a defined value for generating sufficient power\\
\hline
\end{tabular}
\end{table}

 \bibliographystyle{elsarticle-num} 
 \bibliography{bibliography}





% \end{thebibliography}
\end{document}

%% file: acronyms.tex
\makenomenclature
\nomenclature{$VLBI$}{Very Long Baseline Interferometry}
\nomenclature{$EHT$}{Event Horizon Telescope}
\nomenclature{$THEZA$}{TeraHertz Exploration and Zooming-in for \newline Astrophysics}
\nomenclature{$SMBH$}{Super Massive Black Hole}
\nomenclature{$LEO$}{Low Earth Orbit}
\nomenclature{$GEO$}{Geostationary Earth Orbit}
\nomenclature{$HEO$}{Highly Elliptic Orbit}
\nomenclature{$ngEHT$}{Next Generation EHT}
\nomenclature{$GRMHD$}{General Relativistic Magnetohydrodynamics}
\nomenclature{$SNR$}{Signal-Noise Ratio}
\nomenclature{$VSOP$}{VLBI Space Observatory Programme}
\nomenclature{$GR$}{General Relativity}
\nomenclature{$MEO$}{Medium Earth Orbit}
\nomenclature{$EHI$}{Event Horizon Imager}
\nomenclature{$ESA$}{European Space Agency}
\nomenclature{$EHE$}{Event Horizon Explorer}
\nomenclature{$BHEX$}{Black Hole Explorer}
\nomenclature{$SMEX$}{Small Explorer}
\nomenclature{$ISL$}{Inter-Satellite Link}
\nomenclature{$AGN$}{Active Galactic Nuclei}
\nomenclature{$LISA$}{Laser Interferometer Space Antenna}
\nomenclature{$LIGO$}{Laser Interferometer Gravitational-wave Observatory}
\nomenclature{$FOV$}{Field-Of-View}
\nomenclature{$ECI$}{Earth Centered Inertial}
\nomenclature{$STR$}{Star Tracker}
\nomenclature{$DSB$}{Double-Side-Band}
\nomenclature{$SSB$}{Single-Side-Band}
\nomenclature{$USO$}{Ultra-Stable Oscillator}
\nomenclature{$GW$}{Gravitational Wave}

% Add acronyms
\printnomenclature